\documentclass[a4paper, 10pt]{article}
\pdfoutput=1

\usepackage{jheppub}

\usepackage{macros}
\usepackage{dcolumn}
\usepackage{bm}
\usepackage{multirow}
\usepackage{verbatim}
\usepackage{graphicx}
\usepackage[normalem]{ulem}
\usepackage{color}
\usepackage{bbm}
\usepackage{simplewick}
\usepackage{slashed}
\usepackage{tikz}
\usetikzlibrary{shapes.misc}
\usepackage{subfig}
\usepackage{fancyhdr}
\usepackage{feynmp-auto-mod}
\usepackage{simpler-wick}

\usepackage{dsfont}

\newenvironment{myfmf}[1]
{\begin{fmffile}{#1}
\fmfcmd{%
  style_def wboson expr p =
  cdraw (wiggly p);
  shrink (1);
  cfill (arrow p);
  endshrink;
  enddef;}
\fmfcmd{%
  style_def momins expr p =
  drawarrow p;
  enddef;}
\fmfcmd{%
  style_def marrowc expr p =
  drawarrow subpath (1/4, 3/4) of p withpen pencircle scaled 0.4;
  enddef;}
\fmfcmd{%
  style_def marrowd expr p =
  drawarrow subpath (1/4, 3/4) of p shifted 10 down withpen pencircle scaled 0.4;
  enddef;}
\fmfcmd{%
  style_def marrowu expr p =
  drawarrow subpath (1/4, 3/4) of p shifted 10 up withpen pencircle scaled 0.4;
  enddef;}
\fmfcmd{%
  style_def marrowl expr p =
  drawarrow subpath (1/4, 3/4) of p shifted 10 left withpen pencircle scaled 0.4;
  enddef;}
\fmfcmd{%
  style_def mlarrowd expr p =
  drawarrow subpath (1/8, 7/8) of p shifted 10 down withpen pencircle scaled 0.4;
  enddef;}
\fmfcmd{%
  style_def mlarrowu expr p =
  drawarrow subpath (1/8, 7/8) of p shifted 10 up withpen pencircle scaled 0.4;
  enddef;}
\fmfcmd{%
  style_def mlarrowl expr p =
  drawarrow subpath (1/8, 7/8) of p shifted 10 left withpen pencircle scaled 0.4;
  enddef;}
\fmfcmd{%
  style_def marrowr expr p =
  drawarrow subpath (1/4, 3/4) of p shifted 10 right withpen pencircle scaled 0.4;
  enddef;}
\fmfcmd{%
  style_def darrowd expr p =
  drawarrow subpath (1/4, 3/4) of p shifted 10 down dashed evenly withpen pencircle scaled 0.4;
  enddef;}
\fmfcmd{%
  style_def darrowu expr p =
  drawarrow subpath (1/4, 3/4) of p shifted 10 up dashed evenly withpen pencircle scaled 0.4;
  enddef;}
\fmfcmd{%
  style_def darrowl expr p =
  drawarrow subpath (1/4, 3/4) of p shifted 10 left dashed evenly withpen pencircle scaled 0.4;
  enddef;}
\fmfcmd{%
  style_def darrowr expr p =
  drawarrow subpath (1/4, 3/4) of p shifted 10 right dashed evenly withpen pencircle scaled 0.4;
  enddef;}
}
{
\end{fmffile}
}

\newcommand{\msbar}{$\overline{\text{MS}}$}

\renewcommand{\p}{\partial}
\newcommand{\dD}{d^D\!}
\newcommand{\mytag}{\\[-\baselineskip] \stepcounter{equation}\tag{\theequation}}
\renewcommand{\nn}{\nonumber\\}
\renewcommand{\L}{\mathcal{L}}
\renewcommand{\O}{\mathcal{O}}
\newcommand{\M}{\mathcal{M}}
\newcommand{\<}{\langle}
\renewcommand{\>}{\rangle}

\newcommand{\dHV}{\delta_\mathrm{HV}}
\newcommand{\GeV}{\;\text{GeV}}

\allowdisplaybreaks[1]

\begin{document}

\preprint{
\mbox{}\hfill{} JLAB-THY-21-3534 \\
\mbox{}\hfill{} LA-UR-21-31507 \\
\mbox{}\hfill{} PSI-PR-21-28 \\
\mbox{}\hfill{} UWThPh 2021-23
}

\title{One-loop matching for quark dipole operators \\ in a gradient-flow scheme}

\author[a]{Emanuele Mereghetti,}
\emailAdd{emereghetti@lanl.gov}
\affiliation[a]{Theoretical Division, Los Alamos National Laboratory, Los Alamos, NM 87545, USA}

\author[b,c]{Christopher J.~Monahan,}
\emailAdd{cjmonahan@wm.edu}
\affiliation[b]{Department of Physics, The College of William \& Mary, Williamsburg, VA 23187, USA}
\affiliation[c]{Theory Center, Thomas Jefferson National Accelerator Facility, Newport News, VA 23606, USA}

\author[d]{Matthew~D.~Rizik,}
\emailAdd{rizik@nscl.msu.edu}

\author[d]{Andrea Shindler,}
\emailAdd{shindler@frib.msu.edu}
\affiliation[d]{Facility for Rare Isotope Beams \& Physics Department, Michigan State University, \\ East Lansing, Michigan 48824, USA}

\author[e,f,g]{Peter Stoffer}
\emailAdd{stoffer@physik.uzh.ch}
\affiliation[e]{Physik-Institut, Universit\"at Z\"urich, Winterthurerstrasse 190, 8057 Z\"urich, Switzerland}
\affiliation[f]{Paul Scherrer Institut, 5232 Villigen PSI, Switzerland}
\affiliation[g]{University of Vienna, Faculty of Physics, Boltzmanngasse 5, 1090 Vienna, Austria}

\abstract{
The quark chromoelectric dipole (qCEDM) operator is a CP-violating operator
describing, at hadronic energies, beyond-the-standard-model contributions
to the electric dipole moment of particles with nonzero spin.
In this paper we define renormalized dipole operators in a regularization-independent scheme
using the gradient flow, and we perform the matching at one loop in perturbation theory
to renormalized operators of the same
and lower dimension in the more familiar MS scheme.
We also determine the matching coefficients for the 
quark chromo-magnetic dipole operator (qCMDM), which contributes for example to matrix elements
relevant to CP-violating and CP-conserving kaon decays.
The calculation provides a basis for future
lattice QCD computations of hadronic matrix elements of the qCEDM and qCMDM operators.
}

\maketitle

\begin{myfmf}{diags/diags}


\section{Introduction}
\label{sec:intro}

The baryon asymmetry of the universe cannot be explained by known sources of charge 
(C) and parity (P) violation in the standard model of particle 
physics~\cite{Gavela:1993ts,Gavela:1994dt,Gavela:1994ds,Huet:1994jb}. 
CP violation in the standard model occurs through the Cabibbo--Kobayashi--Maskawa (CKM) 
quark-mixing matrix, the Pontecorvo--Maki--Nakagawa--Sakata (PMNS) neutrino mixing matrix, 
and, potentially, could be generated by the quantum chromodyamics (QCD) $\theta$ term. 
Electric dipole moments (EDMs) capture the distribution of positive and 
negative charge within different systems, and the existence of a permanent EDM 
in neutral particles, such as the neutron, can only occur in the presence of CP-violating interactions. 
Experimental constraints on the neutron 
EDM, $d_n = (0.0\pm 1.1_\mathrm{stat}\pm0.2_\mathrm{sys}) \times 10^{-26}$~e~cm at the 90\% 
confidence level~\cite{nEDM:2020crw}, leave open the possibility of unknown sources of 
CP violation several orders of magnitude larger than those generated by the 
standard model~\cite{Shabalin:1979gh,Khriplovich:1981ca,Czarnecki:1997bu,Seng:2014lea}. 
For a full review of EDM searches in a wide range of systems, see Ref.~\cite{Chupp:2017rkp}.

Under the assumption that these unknown sources of CP violation are due to heavy physics beyond the standard model (BSM),
their indirect low-energy effects can be described in terms of effective field theories. Above the electroweak scale, this is the standard model effective field theory (SMEFT), provided that electroweak symmetry is linearly realized~\cite{Buchmuller:1985jz,Grzadkowski:2010es}, while below the weak scale one
should use the low-energy effective field theory (LEFT), which is invariant under the $SU(3)_c \times U(1)_\mathrm{em}$ gauge group. The complete running
and matching in these effective theories up to dimension six has been worked out to one-loop accuracy~\cite{Dekens:2013zca,Alonso:2013hga,Jenkins:2013wua,Jenkins:2013zja,Jenkins:2017dyc,Jenkins:2017jig,Dekens:2019ept}.
At hadronic scales, the CP-violating BSM effects are described by higher-dimensional operators within the LEFT. 
The impact of these operators on the neutron EDM must be disentangled from the standard model 
contributions generated by the CKM matrix and the QCD $\theta$ term. 
One therefore needs to determine individual contributions to the nucleon EDM and to 
CP-odd pion-nucleon and nucleon-nucleon couplings, which
determine nuclear EDM and Schiff moments, at hadronic energy scales. 
Experimental bounds on the neutron EDM are expected to improve by an order of 
magnitude with the next generation of experiments~\cite{Martin:2020lbx}. 
Current determinations of the relevant hadronic matrix elements are primarily 
based on model estimates from 
QCD sum rules~\cite{Pospelov:1999mv,Pospelov:2000bw,Demir:2003js,Pospelov:2005pr,Fuyuto:2012yf,Haisch:2019bml}
or from chiral perturbation theory~\cite{Crewther:1979pi,Ottnad:2009jw,Mereghetti:2010kp,deVries:2010ah,deVries:2012ab}, which 
have large, $\mathcal O(50\%-100\%)$, uncertainties and preclude rigorous 
reduction of systematic uncertainties in theoretical predictions. 
First-principles calculations of the hadronic matrix elements 
with controlled uncertainties are required to exploit fully the 
improved bounds from upcoming experiments. 
Lattice QCD, 
in which QCD is formulated on a Euclidean hypercubic spacetime 
lattice, and correlation functions are determined stochastically, 
provides the best approach for first-principles calculations of QCD at hadronic energies.

Lattice calculations of the matrix elements relevant to the nucleon EDM are challenging for two reasons. 
First, in Euclidean space the QCD $\theta$ term is complex, which prevents the efficient 
use of Monte Carlo methods. This difficulty can be circumvented by expanding the 
Euclidean action in $\theta$, justified by the current experimental bound on the neutron EDM, 
which implies $\theta \sim 10^{-10}$. This approach is theoretically well-defined, 
but faces a challenging signal-to-noise problem associated with the insertion of the $\theta$ 
term in correlation functions on the lattice. This problem can be mitigated through 
signal-to-noise-improved ratios~\cite{Dragos:2018uzd}. Second, renormalization of 
higher-dimensional composite operators on the lattice is difficult. 
For example, the quark chromoelectric dipole moment (qCEDM) operator mixes under renormalization 
with the lower-dimensional pseudoscalar density. On the lattice, 
this mixing is proportional to an inverse power of the lattice spacing and diverges in the continuum limit. 
This power-divergent mixing must be removed nonperturbatively to extract meaningful 
results~\cite{Maiani:1991az}.
For a recent review on lattice-QCD results for the nucleon EDM, see Ref.~\cite{Shindler:2021bcx}.

We proposed applying the gradient flow~\cite{Luscher:2010iy,Luscher:2013cpa} 
to renormalize both the QCD $\theta$ term and the BSM CP-violating operators, 
such as the qCEDM~\cite{Shindler:2014oha,Kim:2018rce}\footnote{For a recent proposal to use 
the gradient flow to smoothen the short-distance behavior of the static potential 
and to improve the approach to the continuum limit see Ref.~\cite{Brambilla:2021egm}.}. 
At finite flow time the qCEDM is multiplicatively renormalizable, 
but to relate this operator to physical matrix elements, 
one must perform an operator-product expansion at short flow times~\cite{Luscher:2013vga}. 
We studied the leading-order short flow-time expansion of the qCEDM and the 
CP-odd three-gluon operator at one-loop in perturbation theory in Ref.~\cite{Rizik:2020naq} 
and nonperturbatively in Ref.~\cite{Kim:2021qae}, 
and our nonperturbative implementation of the hadronic matrix elements required for this program is 
ongoing~\cite{Shindler:2014oha,Shindler:2015aqa,Dragos:2017wms,Dragos:2018uzd,Kim:2018rce,Dragos:2019oxn}. 
Alternative nonperturbative methods for higher-dimensional 
operators are regularization-independent momentum-subtraction schemes, 
which have been defined for the qCEDM~\cite{Bhattacharya:2015rsa} and 
the three-gluon operator~\cite{Cirigliano:2020msr}, 
as well as coordinate-space methods~\cite{Izubuchi:2020ngl}. 
The mixing of these operators under renormalization was studied first 
in~\cite{Weinberg:1989dx,Braaten:1990zt,Braaten:1990gq}, 
then calculated at two loops for the qCEDM in~\cite{Degrassi:2005zd} 
and at two and three loops for the three-gluon operator in~\cite{deVries:2019nsu}.

Power-divergent mixing with lower-dimensional operators hampers the 
renormalization of the quark chromomagnetic dipole moment 
(qCMDM) as well~\cite{Constantinou:2015ela}. 
The flavor-changing qCEDM and qCMDM operators describe low-energy effects of heavy SM and BSM
particles on flavor observables, such as the CP-conserving long-distance contributions to 
$K^0 - \overline{K}^0$ mixing~\cite{DAmbrosio:1999exg}, direct CP-violation in hyperon 
decays~\cite{He:1999bv}, $\epsilon'/\epsilon$ and the $\Delta I=1/2$, 
$K \rightarrow \pi\pi$ transition~\cite{Buras:1999da},
or the CP-violating part of the $K \rightarrow 3 \pi$ decay~\cite{DAmbrosio:1999exg}. Furthermore, 
matrix elements of the flavor-conserving 
qCMDM can be used to extract CP-odd pion-nucleon couplings~\cite{deVries:2012ab,Seng:2016pfd,deVries:2016jox}, which contribute to nuclear Schiff moments. 
A first lattice determination of the matrix element relevant to 
$K^0 - \overline{K}^0$ mixing has been obtained by ETMC using 
twisted mass fermions~\cite{Constantinou:2017sgv}.
Here we propose to use the same strategy adopted for the qCEDM to renormalize the qCMDM operator with the 
gradient flow.

In~\cite{Rizik:2020naq} we determined at one loop in perturbation theory the leading contributions 
to the short flow-time expansion of the qCEDM, which are generated 
by the dimension-three pseudoscalar density operator and the dimension-four topological charge density. 
Here we extend this calculation to determine the dimension-five 
contributions to the short flow-time expansions of the qCEDM 
and the related qCMDM operator. 
We include the complete set of operators up to dimension five that mix 
with qCEDM and qCMDM operators and extract the corresponding short flow-time 
expansion coefficients at one-loop order in perturbation theory. 
These coefficients are necessary to relate the hadronic matrix elements of the 
qCEDM and qCMDM operators, determined nonperturbatively from lattice QCD, 
to their counterparts in the MS scheme, 
which provide inputs into the phenomenological analysis of nucleon EDM measurements.

We organize the rest of this paper as follows. In Sec.~\ref{sec:gf}, 
we introduce the gradient flow and notation relevant for our discussion 
of the short flow-time expansion in Sec.~\ref{sec:sfte}. 
We then determine the matching coefficients to the MS scheme in Sec.~\ref{sec:ren}. 
In Sec.~\ref{sec:scale} we discuss the scale dependence of the matching coefficients, and
we summarize our results and conclusions in Sec.~\ref{sec:summary}.
In the Appendices \ref{sec:conventions} and \ref{sec:FeynmanRules} we list our 
conventions and Feynman rules.


\section{Gradient flow}
\label{sec:gf}

The gradient flow introduces an additional coordinate $t$ of mass dimension $[t]=-2$, 
the so-called flow time (not to be confused with the Minkowski time coordinate---in the following, 
$t$ refers to the flow time)~\cite{Luscher:2010iy,Luscher:2013cpa}. 
Euclidean QCD (see App.~\ref{sec:conventions} and \ref{sec:FeynmanRules} for our conventions) can be regarded as the boundary theory of a $D+1$-dimensional field theory at 
$t=0$~\cite{Luscher:2011bx}. Integrating out suitable Lagrange-multiplier fields in the 
$D+1$-dimensional action is equivalent to imposing the following gradient-flow 
equations on gauge fields, $B_\mu$,  and quark fields, $\chi$, $\chibar$, in $D$ dimensions:
\begin{align}
	\label{eq:FlowEquations}
	\p_t B_\mu &= D_\nu G_{\nu\mu} + \alpha_0 D_\mu \p_\nu B_\nu \, , \nn
	\p_t \chi &= D_\mu D_\mu \chi - \alpha_0 (\p_\mu B_\mu) \chi \, , \nn
	\p_t \bar\chi &= \bar\chi \overleftarrow D_\mu \overleftarrow D_\mu + \alpha_0 \bar \chi \p_\mu B_\mu \, ,
\end{align}
where\footnote{We use the same symbol for the flowed field-strength tensor as for the field-strength tensor at zero flow time.}
\begin{align}
	G_{\mu\nu} = \p_\mu B_\nu - \p_\nu B_\mu + [ B_\mu, B_\nu ] \, ,
\end{align}
together with the boundary conditions
\begin{align}
	B_\mu(x;t=0) &= G_\mu(x) \, , \nn
	\chi(x;t=0) &= \psi(x) \, , \nn
	\bar\chi(x;t=0) &= \bar\psi(x) \, .
\end{align}
Here $\alpha_0$ is a free (gauge) parameter, required for perturbative calculations. 

The (differential) flow equations of~\eqref{eq:FlowEquations}, together with boundary conditions, can be rewritten as integral equations~\cite{Luscher:2011bx}
\begin{align}
	\label{eq:FlowIntegralEquations}
	B_\mu(x;t) &= \int \dD y \left[ K_{\mu\nu}(x-y;t) G_\nu(y) + \int_0^t ds K_{\mu\nu}(x-y;t-s) R_\nu(y;s) \right] \, , \nn
	\chi(x;t) &= \int \dD y \left[ J(x-y;t) \psi(y) + \int_0^t ds J(x-y;t-s) \Delta'\chi(y;s) \right] \, , \nn
	\bar\chi(x;t) &= \int \dD y \left[ \bar\psi(y) \bar J(x-y;t) + \int_0^t ds \bar\chi(y;s) \overleftarrow\Delta' \bar J(x-y;t-s) \right] \, ,
\end{align}
where the heat kernels are
\begin{align}
	K_{\mu\nu}(x;t) &= \int \frac{\dD p}{(2\pi)^D} \frac{e^{ip\cdot x}}{p^2} \left[ (\delta_{\mu\nu}p^2 - p_\mu p_\nu) e^{-tp^2} + p_\mu p_\nu e^{-\alpha_0 t p^2} \right] \, , \nn
	J(x;t) &= \bar J(x;t) = \int \frac{\dD p}{(2\pi)^D} e^{ip\cdot x} e^{-t p^2} = \frac{e^{-\frac{x^2}{4t}}}{(4\pi t)^{D/2}} \, ,
\end{align}
and the interaction terms are given by
\begin{align}
	\label{eq:FlowInteractionTerms}
	R_\mu &= 2 [ B_\nu, \p_\nu B_\mu] - [B_\nu, \p_\mu B_\nu] + (\alpha_0-1) [B_\mu, \p_\nu B_\nu] + [B_\nu, [B_\nu, B_\mu]] \, , \nn
	\Delta' &= (1-\alpha_0) (\p_\nu B_\nu) + 2 B_\nu \p_\nu + B_\nu B_\nu \, , \nn
	\overleftarrow \Delta' &= -(1-\alpha_0) (\p_\nu B_\nu) - 2 \overleftarrow \p_\nu B_\nu + B_\nu B_\nu \, .
\end{align}
The flow equations in integral form~\eqref{eq:FlowIntegralEquations} can be solved iteratively, which corresponds to an expansion in powers of $g_0$ upon rescaling the gauge field $B_\mu \mapsto g_0 B_\mu$. This allows one to express the flowed (bulk) fields in terms of the fundamental fields at the boundary $t=0$. From the expansion of the kernel, one obtains propagator-like structures called flow lines. The interaction terms in~\eqref{eq:FlowInteractionTerms} induce interaction vertices with three and four fields, the flow vertices. Our conventions for the Feynman rules and diagrams are given in App.~\ref{sec:FeynmanRules}.


\section{Short flow-time expansion}
\label{sec:sfte}

In the following, we consider Green's functions with operator insertions at finite flow time $t$. 
The goal is to extract the relation between renormalized flowed operators and MS operators at 
zero flow time:
\begin{align}
	\label{eq:ShortFlowtimeOPE}
	\O_i^R(t) = \sum_j c_{ij}(t,\mu) \, \O_j^{\textrm{MS}}(\mu) \, .
\end{align}
This ``short flow-time expansion'' (SFTE) is an operator-product expansion (OPE) that is valid at small flow time $t$, 
where the hard scale is proportional to $t^{-1/2}$.
We will take into account operators up to dimension five  
on both sides of the matching equation~\eqref{eq:ShortFlowtimeOPE}.

To extract the flow-time dependent coefficients $c_{ij}(t,\mu)$, we consider 
insertions of the operators $\O_i^R(t)$ in off-shell amputated one-particle irreducible (1PI) 
Green's functions. We work in the massless limit and consider the matching to one-loop accuracy.

The relation between amputated Green's functions of the bare operators and 
renormalized Green's functions is schematically given by
\begin{align}
	 \left\langle \left(\psi^{(0)}\right)^{n_\psi}  \right. & \left. \left(\psibar^{(0)}\right)^{n_{\psibar}} \left(G_\mu^{(0)}\right)^{n_G} \O_i^{(0)}[\psi^{(0)}, \psibar^{(0)}, G^{(0)}] \right\rangle^\text{amp} \nn 
	&= Z_\psi^{-(n_\psi + n_{\psibar})/2} Z_G^{-n_G/2} Z_{ij}^\text{MS} \left\langle \left(\psi\right)^{n_\psi} \left(\psibar\right)^{n_{\psibar}} \left(G_\mu\right)^{n_G}\O_j^\text{MS}[\psi, \psibar, G] \right\rangle^\text{amp} \nn
		&= Z_\psi^{-(n_\psi + n_{\psibar})/2} Z_G^{-n_G/2} Z_{ij}^\text{MS} c^{-1}_{jk} \left\langle \left(\psi\right)^{n_\psi} \left(\psibar\right)^{n_{\psibar}} \left(G_\mu\right)^{n_G} \O_k^R[\chi, \chibar, B] \right\rangle^\text{amp} \nn
		&= Z_{ij}^\text{MS} c^{-1}_{jk} Z_\chi^{-n/2} \left\langle \left(\psi^{(0)}\right)^{n_\psi} \left(\psibar^{(0)}\right)^{n_{\psibar}} \left(G_\mu^{(0)}\right)^{n_G}\O_k^t[\chi^{(0)}, \chibar^{(0)}, B^{(0)}] \right\rangle^\text{amp} \, ,
\end{align}
where we allow a generic number of $n_\psi$ external fermion, $n_{\psibar}$ antifermion, and $n_G$ gauge fields at zero flow time.\footnote{In Sec.~\ref{sec:qCEDM} we also use an external photon field to study the mixing 
of the quark chromo-EDM with the quark EDM operator, though the photon renormalization constant $Z_A$ does not contribute at leading order.} Using standard procedures 
we renormalize each field with the corresponding $Z^{-1/2}$ renormalization factor
and denote with $Z_{ij}^\text{MS}$ the matrix renormalizing $\O_i^{(0)}$ in the MS scheme.
The renormalization of the bare flowed operators $\O_k^t$ 
is diagonal and only requires the renormalization of the bare parameters of the QCD theory (coupling and quark masses)
and the flowed quark fields
with a factor $Z_\chi^{-n/2}$, where $n$ denotes the total number of fermion and antifermion fields
in the operator $\O_k^t$.
Since the external states are at zero flow time, 
the product of the wave-function renormalization factors
cancels in the matching equation:\footnote{The external-leg amputation happens at zero flow time, which can leave an exponential factor as a remainder if the external legs connect to a vertex at finite flow time.}
\begin{align}
	\label{eq:MatchingEquation}
	c_{ij} \, \left(Z_{jk}^\text{MS}\right)^{-1}\Big\< \left(\psi^{(0)}\right)^{n_\psi} & \left(\psibar^{(0)}\right)^{n_{\psibar}} \left(G_\mu^{(0)}\right)^{n_G} \O_k^{(0)}[\psi^{(0)}, \psibar^{(0)}, G^{(0)}] \Big\>^\text{amp} \nn
		&= Z_\chi^{-n/2} \Big\<  \left(\psi^{(0)}\right)^{n_\psi} \left(\psibar^{(0)}\right)^{n_{\psibar}} \left(G_\mu^{(0)}\right)^{n_G}\O_i^t[\chi^{(0)}, \chibar^{(0)}, B^{(0)}] \Big\>^\text{amp} \, .
\end{align}


\section{Renormalization and matching coefficients}
\label{sec:ren}

\subsection{Dirac algebra}

We perform the calculation both in the 't\,Hooft--Veltman (HV) scheme~\cite{tHooft:1972tcz,Breitenlohner:1977hr} as well as in the scheme with anticommuting $\gamma_5$ (NDR). For the CP-odd operators, we define the following Dirac structures:
\begin{align}
	\tilde\sigma_{\mu\nu}^\mathrm{HV} &= -\frac{1}{2} \epsilon_{\mu\nu\alpha\beta} \sigma_{\alpha\beta} \, , \quad
	\tilde\sigma_{\mu\nu}^\mathrm{NDR} = \sigma_{\mu\nu}\gamma_5 \, ,
\end{align}
where as usual $\sigma_{\mu\nu} = \frac{i}{2} [\gamma_\mu, \gamma_\nu ]$. As the Levi-Civita symbol is a purely four-dimensional object, in the HV scheme $\tilde\sigma_{\mu\nu}$ only contains four-dimensional components. In order to compare the HV and NDR schemes, we introduce
\begin{align}
	\dHV = \left\{ \begin{matrix} 1 & \text{ in the HV scheme,} \\ 0 & \text{ in the NDR scheme.} \end{matrix} \right.
\end{align}

\subsection{Quark-field renormalization}

We express the SFTE in terms of operators in the MS scheme. Therefore, we renormalize the parameters of the boundary theory in the MS scheme and define the renormalized coupling as
\begin{align}
	g_0 = Z_g g \mu^\varepsilon \, .
\end{align}
One can easily switch to the \msbar{} scheme by replacing the MS scale $\mu$ with the \msbar{} scale $\bar\mu$ according to
\begin{align}
	\mu = \bar\mu \frac{e^{\gamma_E/2}}{(4\pi)^{1/2}} \, .
\end{align}

While the flow equations regulate most of the UV singularities, the fermion fields (as well as the gauge coupling and quark mass) require renormalization. In the MS scheme, we define renormalized flowed fermion fields according to
\begin{align}
	\chi^{(0)}(x;t) = Z_\chi^{1/2} \chi(x;t) \, , \quad \bar\chi^{(0)}(x;t) = Z_\chi^{1/2} \bar\chi(x;t) \, ,
\end{align}
where the superscript ${(0)}$ marks the bare fields and
\begin{align}
	\label{eq:QuarkFieldsRenormalization}
	Z_\chi = 1 - \frac{\alpha_s C_F}{4\pi} \frac{3}{\varepsilon} \, .
\end{align}
The quark-field renormalization leads to a counterterm contribution to the two-point function of renormalized quark fields of $\tilde S_\text{ct}(p,s,t) = (Z_\chi^{-1} - 1) \tilde S(p,s+t)$, where the propagator is defined in~\eqref{eq:QuarkPropagator}. Summing tree-level and one-loop diagrams leads to the following finite result for the two-point function in the massless limit:
\begin{align}
	\tilde S_{\text{NLO}}(p,s,t) &= \int d^Dx e^{-i p \cdot x} \< \chi(x;t) \bar \chi(0;s) \> \Big|_{\text{NLO}} \\
		&= \tilde S(p,s+t) \biggl( 1 - \frac{\alpha_s C_F}{4\pi} \begin{aligned}[t]
		&\biggl[ \xi \log\left( \frac{4\pi\mu^2}{p^2}\right) + 1 - \xi \gamma_E - \xi\log(\alpha_0) \nn
		&+ \frac{3-\xi}{2} \log(8\pi\mu^2t) + \frac{3-\xi}{2} \log(8\pi\mu^2s) + \O(p^2t,p^2s)\biggr] \biggr) \, . \end{aligned}
\end{align}
Note that the finite part of the two-point function depends on the gauge parameters, while the flowed quark-field renormalization $Z_\chi$ in~\eqref{eq:QuarkFieldsRenormalization} is independent of $\xi$ and $\alpha_0$.

To make contact with lattice calculations, it is necessary to implement a renormalization scheme that is regularization independent. This is achieved by imposing the following regularization-independent renormalization condition on a gauge-invariant composite operator (for one quark flavor)~\cite{Makino:2014wca,Makino:2014taa}
\begin{align}
	\label{eq:vevRenormalizationCondition}
	\< 0 | \mathring{\bar\chi}(x;t) \overleftrightarrow {\slashed D} \mathring\chi(x;t) | 0 \> = - \frac{2 N_c}{(4\pi)^2 t^2} \, ,
\end{align}
where $\overleftrightarrow D_\mu = D_\mu - \overleftarrow D_\mu$. In dimensional regularization, this implies an additional finite renormalization compared to MS. The ``ringed fields'' are related to the MS renormalized fields by
\begin{align}
	\chi(x;t) = (8\pi t)^{\varepsilon/2} \zeta_\chi^{1/2} \mathring\chi(x;t) \, , \quad \bar\chi(x;t) = (8\pi t)^{\varepsilon/2} \zeta_\chi^{1/2} \mathring{\bar\chi}(x;t) \, .
\end{align}
The prefactors $(8\pi t)^{\varepsilon/2}$ are introduced in dimensional regularization because the renormalization condition fixes the dimension of the fields $\mathring{\bar\chi}$ and $\mathring\chi$ to be equal to $3/2$ instead of $(D-1)/2$.

The next-to-leading order (NLO) contribution to the vacuum expectation value in~\eqref{eq:vevRenormalizationCondition} is obtained from vacuum two-loop diagrams~\cite{Makino:2014taa}, leading to the following finite renormalization $\zeta_\chi$:
\begin{align}
	\label{eq:RingedFieldsRenormalization}
	\zeta_\chi = 1 - \frac{\alpha_s C_F}{4\pi} \left( 3 \log\left( 8\pi \mu^2 t \right) - \log(432) \right) \, .
\end{align}
We have performed the calculation of $\zeta_\chi$ for generic $\xi$ and $\alpha_0$, confirming its gauge-parameter independence.

\subsection{Expanding loops}
\label{sec:ExpandingLoops}

The matching coefficients $c_{ij}(t,\mu)$ only depend on the flow-time $t$ and the MS renormalization scale $\mu$ and can be expanded in the renormalized MS coupling $\alpha_s = g^2/(4\pi)$ as
\begin{align}
	c_{ij}(t,\mu) = \delta_{ij} + \frac{\alpha_s(\mu)}{4\pi} c_{ij}^{(1)}(t,\mu) + \O(\alpha_s^2) \, .
\end{align}
The coefficients $c_{ij}^{(1)}$ are independent of the soft scales---we include powers of the quark mass explicitly in the operators. When solving the matching equation for the coefficients $c_{ij}^{(1)}$, the non-analytic dependence on the soft scales cancels between the MS and flowed loop diagrams. Therefore, one can apply standard techniques for matching calculations~\cite{Manohar:1996cq,Manohar:1997qy,Manohar:2018aog} and expand the integrands of the loop integrals in all scales apart from the flow time $t$, before integration: although this alters the analytic structure of the loop integrals and distorts the infrared (IR) structure, these IR modifications drop out in the difference between MS and flowed loop diagrams. Expanding the loop integrals leads to scaleless integrals for the operator insertions on the LHS of~\eqref{eq:MatchingEquation}, which vanish in dimensional regularization, hence ultraviolet (UV) and IR singularities of the expanded loops are identical. Insertions of the flowed operators are free from UV singularities (apart from the renormalization of the gauge coupling and the quark-field renormalization $Z_\chi$). The IR singularities of the expanded loop integrals on the RHS of~\eqref{eq:MatchingEquation} exactly match the UV MS counterterms. The finite matching coefficients $c_{ij}$ can then be most easily obtained from the expanded integrals of insertions of the flowed operators, which are single-scale integrals and straightforward to calculate. Even the inclusion of generic gauge parameters $\xi$ and $\alpha_0$ does not lead to major complications in the calculation of the integrals, hence we perform all calculations for generic $\xi$ and $\alpha_0$, which provides a useful check: the coefficients $c_{ij}$ of gauge-invariant operators in the SFTE need to be independent of $\xi$ and $\alpha_0$.

When expanding the loop integrands in all soft scales before performing the loop integrals, one potential pitfall arises in the calculation due to the fact that the ringed fields are renormalized through the condition~\eqref{eq:vevRenormalizationCondition} and not in the MS scheme, as we will explain in the following.

In the dimensionally regularized theory, we can relate both the renormalized MS and flowed operators to the bare operators at zero flow time by
\begin{align}
	\label{eq:BareVsRenormalizedOps}
	\O^{(0)}_i &= Z_{ik}^\mathrm{MS}(\mu) \O^\mathrm{MS}_k(\mu) = \tilde\mu^{-n\varepsilon} Z_{ik}^R(t,\tilde\mu) \O^R_k(t) \, ,
\end{align}
where the relation for the flowed operators involves the short flow-time OPE and hence an infinite sum and where we assumed for notational simplicity that the operators contain in total $n$ fermion and antifermion fields. We have introduced an arbitrary mass scale $\tilde\mu$, which compensates the mismatch of mass dimension between the bare operators and the flowed operators in terms of ringed fields. If we choose $\tilde\mu = (8\pi t)^{-1/2}$, at one loop the renormalization factors $Z_{ik}^R$ can be written as
\begin{align}
	Z_{ik}^R(t,(8\pi t)^{-1/2}) = \delta_{ik} + \frac{\alpha_s}{4\pi} \left( \frac{\Delta_{ik}}{\varepsilon} + \Delta_{ik}^R \right) + \O(\alpha_s \varepsilon, \alpha_s^2) \, .
\end{align}
For a generic choice of $\tilde\mu$, this changes to
\begin{align}
	Z_{ik}^R(t,\tilde\mu) = \delta_{ik} \left[1 + \frac{n \varepsilon}{2} \log\left( 8\pi \tilde\mu^2 t \right) \right] + \frac{\alpha_s}{4\pi} \left( \frac{\Delta_{ik}}{\varepsilon}  + \frac{\Delta_{ik} n}{2} \log\left( 8\pi\tilde\mu^2 t \right) + \Delta_{ik}^R \right) + \O(\varepsilon^2,\alpha_s \varepsilon, \alpha_s^2) \, ,
\end{align}
where we have included the $\O(\alpha_s^0)$ evanescent structure linear in $\varepsilon$. The matching coefficients follow from~\eqref{eq:BareVsRenormalizedOps}:
\begin{align}
	\O^R_i(t) = \tilde\mu^{n\varepsilon} \left[ (Z^R)^{-1}_{ij}(t,\tilde\mu) Z^\mathrm{MS}_{jk}(\mu) \right] \O^\mathrm{MS}_k(\mu) = \tilde\mu^{n\varepsilon} c_{ik}(t,\mu,\tilde\mu) \O^\mathrm{MS}_k(\mu) \, ,
\end{align}
which is the continuation of~\eqref{eq:ShortFlowtimeOPE} to $D$ dimensions. The coefficient $c_{ik}(t,\mu,\tilde\mu)$ depends on $\tilde\mu$ through evanescent terms. We are only interested in the limit $\epsilon\to0$:
\begin{align}
	c_{ik}(t,\mu) &= \lim_{\varepsilon\to0} c_{ik}(t,\mu,\tilde\mu) = \lim_{\varepsilon\to0} \left[ (Z^R)^{-1}_{ij}(t,\tilde\mu) Z^\mathrm{MS}_{jk}(\mu) \right] \nn
		&= \lim_{\varepsilon\to0} \bigg\{ \left( \delta_{ij} \left[1 - \frac{n \varepsilon}{2} \log\left( 8\pi \tilde\mu^2 t \right) \right] - \frac{\alpha_s}{4\pi} \left( \frac{\Delta_{ij}}{\varepsilon} - \frac{\Delta_{ij} n}{2} \log\left( 8\pi \tilde\mu^2 t \right) + \Delta_{ij}^R \right) \right) \nn
			&\qquad\qquad \times \left[ \delta_{jk} + \frac{\alpha_s}{4\pi} \frac{\Delta_{jk}}{\varepsilon} \right] \bigg\} + \O(\alpha_s^2) \nn
		&= \lim_{\varepsilon\to0} \bigg\{ \delta_{ik} \left[1 - \frac{n \varepsilon}{2} \log\left( 8\pi \tilde\mu^2 t \right) \right] - \frac{\alpha_s}{4\pi} \Delta_{ik}^R  \bigg\} + \O(\alpha_s^2) \nn
		&= \delta_{ik} - \frac{\alpha_s}{4\pi} \Delta_{ik}^R + \O(\alpha_s^2) = (8\pi\tilde\mu^2 t)^{n\varepsilon/2} c_{ik}(t,\mu,\tilde\mu) \, ,
\end{align}
which is indeed independent of $\tilde\mu$, but only if the linearly evanescent $\O(\alpha_s^0)$ term is correctly taken into account.

Including the finite renormalization and the factor $(8\pi t)^{\varepsilon/2}$ that compensates the mass dimension of the ringed fields in the dimensionally regularized theory, the matching equation~\eqref{eq:MatchingEquation} reads
\begin{align}
	&c_{ij}(t,\mu,\tilde\mu) \, \left(Z_{jk}^\text{MS}\right)^{-1} Z_\psi^{(n_\psi + n_{\psibar})/2} Z_G^{n_G/2} \Big\< \left(\psi^{(0)}\right)^{n_\psi} \left(\psibar^{(0)}\right)^{n_{\psibar}} \left(G_\mu^{(0)}\right)^{n_G} \O_k^{(0)}[\psi^{(0)}, \psibar^{(0)}, G^{(0)}] \Big\>^\text{amp} \nn
		&\qquad = (8\pi \tilde\mu^2 t)^{-n \varepsilon/2} Z_\psi^{(n_\psi + n_{\psibar})/2} Z_G^{n_G/2} Z_\chi^{-n/2} \zeta_\chi^{-n/2} \nn
			&\qquad\qquad \times \Big\<  \left(\psi^{(0)}\right)^{n_\psi} \left(\psibar^{(0)}\right)^{n_{\psibar}} \left(G_\mu^{(0)}\right)^{n_G}\O_i^t[\chi^{(0)}, \chibar^{(0)}, B^{(0)}] \Big\>^\text{amp} \, .
\end{align}
We have kept all renormalization factors in the equation, so that both sides are UV finite. In particular, the factor $(8\pi \tilde\mu^2 t)^{-n \varepsilon/2}$ multiplies a UV finite quantity. Its evanescent component matches the evanescent term in $c_{ij}(t,\mu,\tilde\mu)$. Therefore, the matching equation can be simplified to
\begin{align}
	&c_{ij}(t,\mu) \, \left(Z_{jk}^\text{MS}\right)^{-1} Z_\psi^{(n_\psi + n_{\psibar})/2} Z_G^{n_G/2} \Big\< \left(\psi^{(0)}\right)^{n_\psi} \left(\psibar^{(0)}\right)^{n_{\psibar}} \left(G_\mu^{(0)}\right)^{n_G} \O_k^{(0)}[\psi^{(0)}, \psibar^{(0)}, G^{(0)}] \Big\>^\text{amp} \nn
		&\qquad = Z_\psi^{(n_\psi + n_{\psibar})/2} Z_G^{n_G/2} Z_\chi^{-n/2} \zeta_\chi^{-n/2} \Big\<  \left(\psi^{(0)}\right)^{n_\psi} \left(\psibar^{(0)}\right)^{n_{\psibar}} \left(G_\mu^{(0)}\right)^{n_G}\O_i^t[\chi^{(0)}, \chibar^{(0)}, B^{(0)}] \Big\>^\text{amp} \, .
\end{align}
At this stage, the integrands on both side of the equation can be expanded in the soft scales before performing the loop integrals. The loops on the left-hand side are transformed into scaleless integrals, where UV and IR divergences are identical. The same IR divergences are generated on both sides of the matching equation, so that they cancel when solving for $c_{ij}$. However, since after the expansion both sides of the equation contain divergences, it is important that evanescent terms on both sides are consistently taken into account and that the factor $(8\pi \tilde\mu^2 t)^{-n \varepsilon/2}$ is cancelled by the evanescent part of $c_{ij}(t,\mu,\tilde\mu)$.  Indeed, we have confirmed explicitly that all IR poles do cancel on either side by resumming to all orders the contributions from the relevant soft scales.

\subsection{Chromo-EDM}
\label{sec:qCEDM}

We define the flowed qCEDM operator in terms of renormalized (ringed) fields as
\begin{align}
	\label{eq:qCEDMFlowed}
	\O_{CE}^R(x;t) = \mathring{\bar\chi}(x;t) \tilde\sigma_{\mu\nu} t^a \mathring\chi(x;t) G_{\mu\nu}^a(x;t) \, .
\end{align}
We want to extract the coefficients of the short flow-time OPE up to dimension five:
\begin{align}
	\O_{CE}^R(x;t) &= c_P(t,\mu) \O_P^\mathrm{MS}(x;\mu) + c_{m^2P}(t,\mu) \O_{m^2P}^\mathrm{MS}(x;\mu) + c_{m\theta}(t,\mu) \O_{m\theta}^\mathrm{MS}(x;\mu) \nn
		&\quad + c_E(t,\mu) \O_E^\mathrm{MS}(x;\mu) + c_{CE}(t,\mu) \O_{CE}^\mathrm{MS}(x;\mu) + \ldots \, ,
\end{align}
where the MS operators are minimally subtracted versions of the following operators:
\begin{align}
	\O_P(x;\mu) &= \bar\psi(x) \gamma_5 \psi(x) \, , \nn
	\O_{m^2 P}(x;\mu) &= m^2 \bar\psi(x) \gamma_5 \psi(x) \, , \nn
	\O_{m\theta}(x;\mu) &= \mu^{-2\varepsilon} m \, \tr[ G_{\mu\nu} \widetilde G_{\mu\nu} ] \, , \nn
	\O_E(x;\mu) &= \bar\psi(x) \tilde\sigma_{\mu\nu} F_{\mu\nu}(x) \psi(x)  \, , \nn
	\O_{CE}(x;\mu) &= \bar\psi(x) \tilde\sigma_{\mu\nu} t^a \psi(x) G_{\mu\nu}^a(x) \, ,
\end{align}
in terms of renormalized fields with $F_{\mu\nu}$ the field-strength tensor of the external $U(1)$ gauge field. The dual gluonic field-strength tensor is defined as $\widetilde G_{\mu\nu} = \frac{1}{2} \epsilon_{\mu\nu\rho\sigma} G_{\rho\sigma}$. The five coefficients can be extracted by computing insertions of the flowed operator into suitable 1PI Green's functions (including wave-function renormalization).

The mixing with the pseudoscalar density is obtained from the 1PI matrix element with external quark-antiquark states (we can also insert momentum $-q$ into the operator in order to obtain the mixing with total-derivative operators):
\begin{align}
	\int d^Dx e^{-i q \cdot x} \< \psi(k) | \O_{CE}^R(x;t) | \psi(p) \> \big|_\text{amp} 
		&= (2\pi)^D \delta^{(D)}(p-k-q) \< \psi(k) | \O_{CE}^R(0;t) | \psi(p) \> \big|_\text{amp} \nn
		&=: (2\pi)^D \delta^{(D)}(p-k-q) \M(p,k) \, .
\end{align}
In total, one obtains~\cite{Rizik:2020naq}
\begin{align}
	c_P(t,\mu) &= \frac{\alpha_s C_F}{4\pi} \frac{6i}{t} \, , \nn 
	c_{m^2 P}(t,\mu) &   = \frac{\alpha_s C_F}{4\pi} i \left[  12 \log(8\pi\mu^2 t) + \frac{1}{2} (29 + 24 \delta_{\rm HV}) \right] \, ,
\end{align}
where $c_{m^2 P}$ is a new result.

Due to chiral symmetry (for a discussion of chiral symmetry at finite flow time see Ref.~\cite{Shindler:2013bia}),
mixing with the topological charge density requires an insertion of a mass factor, which we include in the operator $\O_{m\theta}$. 
The SFTE coefficient can be extracted by calculating the Green's function
\begin{align}
	\int d^Dx e^{-i q \cdot x} \< g(k) | \O_{CE}^R(x;t) | g(p) \> \big|_\text{amp}
		&= (2\pi)^D \delta^{(D)}(p-k-q) \< g(k) | \O_{CE}^R(0;t) | g(p) \> \big|_\text{amp} \nn
		&=: (2\pi)^D \delta^{(D)}(p-k-q) \epsilon_\mu^a(p) \epsilon_\nu^b(k)^* \M_{\mu\nu}^{ab}(p,k) \, ,
\end{align}
where momentum $-q$ is necessarily inserted into the operator~\cite{Georgi:1980cn}. The loop calculation leads to $\gamma_5$-odd Dirac traces that are not well-defined in NDR. The result in the HV scheme reads~\cite{Rizik:2020naq}
\begin{align}
	c_{m\theta}(t,\mu) = \frac{i}{4\pi^2} \left[ 1 + \log(8\pi\mu^2t) \right] \, .
\end{align}

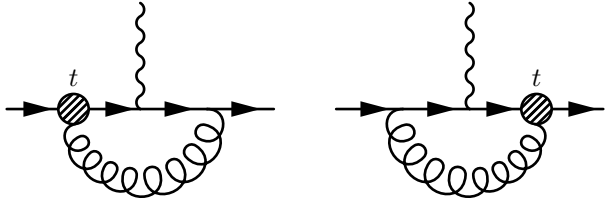
\begin{figure}[t]
	\centering
		\begin{fmfgraph*}(100,80)
			\fmfleft{l1} \fmfright{r1} \fmftop{t1}
			\fmf{quark}{l1,v2,v1,v3,r1}
			\fmffreeze
			\fmf{photon}{v1,t1}
			\fmf{gluon,right}{v2,v3}
			\fmfv{decor.shape=circle, decor.filled=shaded, decor.size=(4mm),label.angle=90,label.dist=3mm,label=$t$}{v2}
		\end{fmfgraph*}
		\qquad
		\begin{fmfgraph*}(100,80)
			\fmfleft{l1} \fmfright{r1} \fmftop{t1}
			\fmf{quark}{l1,v2,v1,v3,r1}
			\fmffreeze
			\fmf{photon}{v1,t1}
			\fmf{gluon,right}{v2,v3}
			\fmfv{decor.shape=circle, decor.filled=shaded, decor.size=(4mm),label.angle=90,label.dist=3mm,label=$t$}{v3}
		\end{fmfgraph*}
	\caption{Feynman diagrams for the matching calculation of the flowed qCEDM operator to the qEDM operator. The hatched blob denotes the insertion of the flowed qCEDM operator at flow time $t$.}
	\label{img:EDMDiagrams}
\end{figure}

\begin{figure}[t]
	\centering
	\includegraphics[width=3cm]{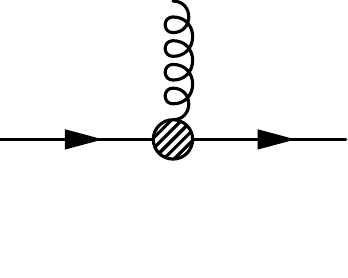} \quad
	\includegraphics[width=3cm]{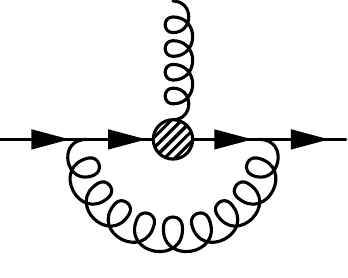} \quad
	\includegraphics[width=3cm]{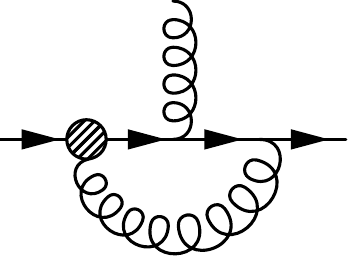} \quad
	\includegraphics[width=3cm]{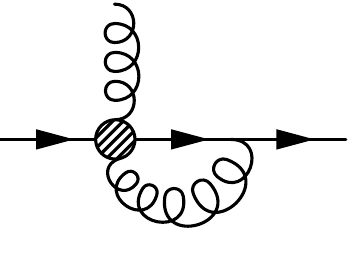} \quad
	\includegraphics[width=3cm]{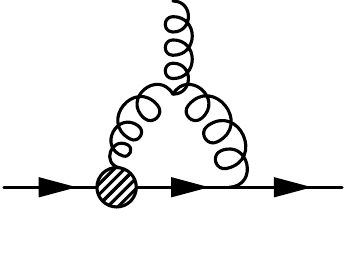} \quad
	\includegraphics[width=3cm]{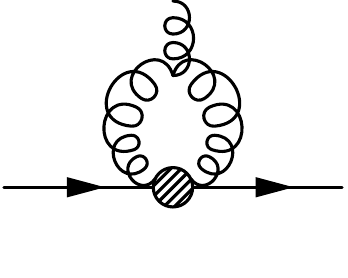} \quad
	\includegraphics[width=3cm]{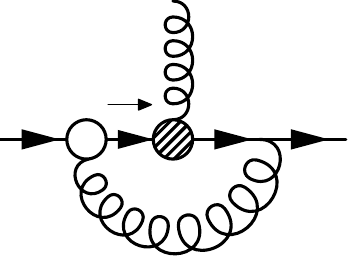} \quad
	\includegraphics[width=3cm]{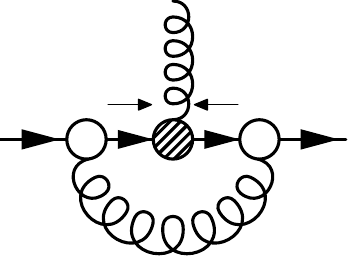} \quad
	\includegraphics[width=3cm]{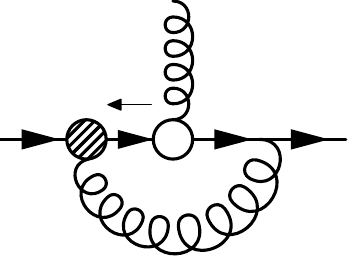} \quad
	\includegraphics[width=3cm]{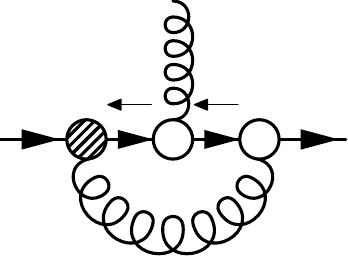} \quad
	\includegraphics[width=3cm]{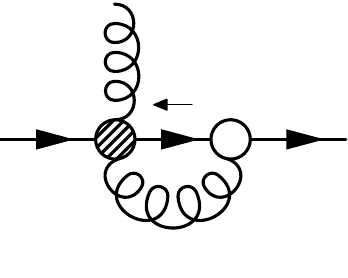} \quad
	\includegraphics[width=3cm]{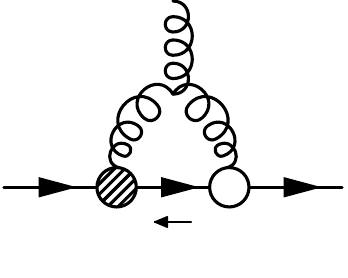} \quad
	\includegraphics[width=3cm]{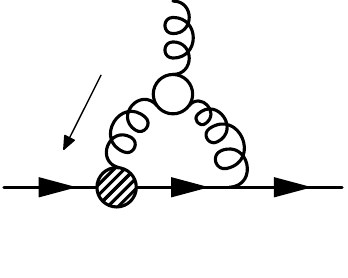} \quad
	\includegraphics[width=3cm]{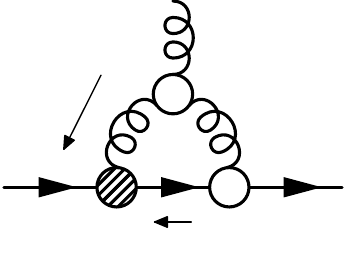} \quad
	\includegraphics[width=3cm]{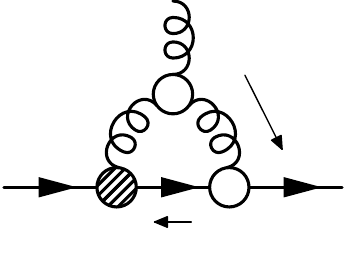} \quad
	\includegraphics[width=3cm]{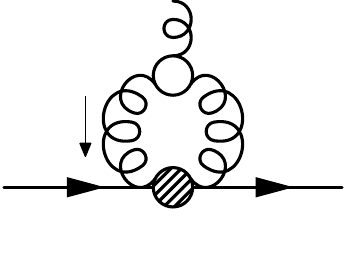} \quad
	\includegraphics[width=3cm]{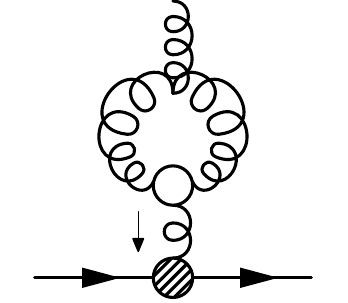} \quad
	\includegraphics[width=3cm]{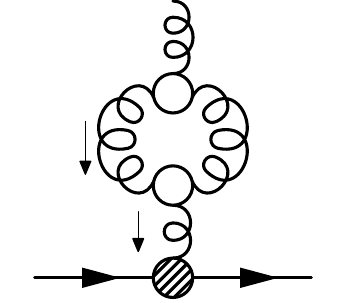} \quad
	\includegraphics[width=3cm]{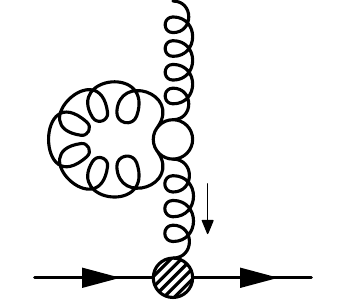} \quad
	\includegraphics[width=3cm]{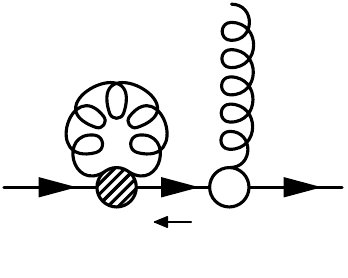} \quad
	\includegraphics[width=3cm]{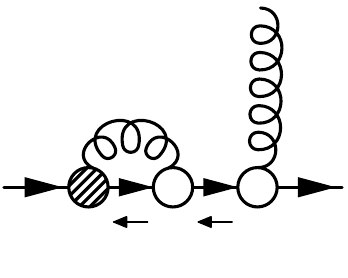} \quad
	\includegraphics[width=3cm]{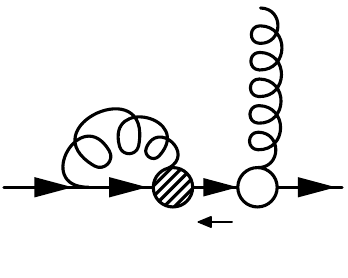} \quad
	\includegraphics[width=3cm]{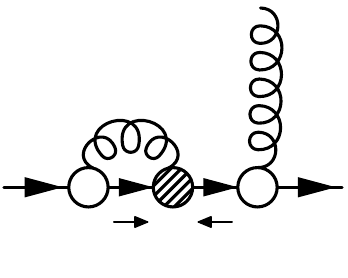} \quad
	\includegraphics[width=3cm]{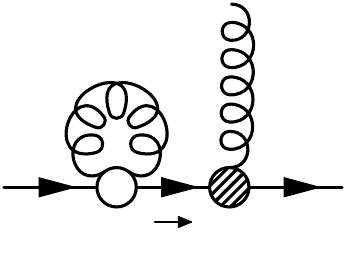} \quad
	\includegraphics[width=3cm]{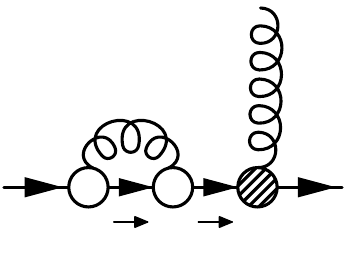} \quad
	\includegraphics[width=3cm]{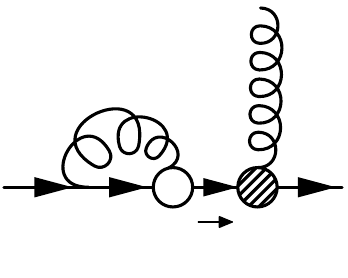} \quad
	\includegraphics[width=3cm]{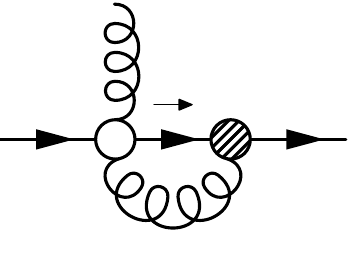}

	\caption{Feynman diagrams for the matching calculation of the flowed qCEDM operator. We do not show 19 additional diagrams that follow from crossing and inverting the fermion-flow direction. The hatched blob denotes the insertion of the flowed qCEDM operator at flow time $t$, while the open circles are flow vertices. Normal quark and gluon lines denote propagators, while lines with an adjacent arrow stand for flow lines. The adjacent arrow points into the direction of increasing flow time.}
	\label{img:cEDMDiagrams}
\end{figure}

In order to extract the mixing with the qEDM operator, we consider a quark-antiquark matrix element with an external $U(1)$ field:
\begin{align}
	\int d^Dx e^{-i q \cdot x} & \< \psi(k) | \O_{CE}^R(x;t) | \psi(p) \gamma(r) \> \big|_\text{amp} \nn
		&= (2\pi)^D \delta^{(D)}(p+r-k-q) \< \psi(k) | \O_{CE}^R(0;t) | \psi(p) \gamma(r) \> \big|_\text{amp} \nn
		&=: (2\pi)^D \delta^{(D)}(p+r-k-q) \epsilon_\mu(r) \M_\mu(p,k,r) \, .
\end{align}
Only two diagrams contribute, shown in Fig.~\ref{img:EDMDiagrams}. As we are not interested in total derivative operators, we can send $q\to0$ and obtain for the diagrams expanded in all soft scales:
\begin{align}
	\M_\mu(p,k,r) &= \frac{\alpha_s C_F}{4\pi} \bigg[ \left( \frac{1}{\varepsilon} + \log(8\pi\mu^2 t) \right) ( 6 \gamma_5 (k_\mu + p_\mu) - 2 i \tilde\sigma_{\mu\nu} r_\nu) \nn
		&\qquad\qquad + \left(\frac{1}{2}+6\delta_\mathrm{HV}\right) \gamma_5 (k_\mu + p_\mu) - \left(\frac{11}{2}-2\delta_\mathrm{HV}\right) i \tilde\sigma_{\mu\nu} r_\nu \bigg] \, .
\end{align}
Up to the contribution of equation-of-motion operators, this results in
\begin{align}
	c_E(t,\mu) = \frac{\alpha_s C_F}{4\pi} \left[ 4 \log(8\pi\mu^2 t) + 3 + 2 \delta_\mathrm{HV} \right] \, .
\end{align}
There is no tree-level contribution to this coefficient, so at the one-loop level it is immaterial whether the flowed fermion fields are renormalized in MS or through the ringed fermion renormalization condition in Eq.~\eqref{eq:vevRenormalizationCondition}.

Finally, the coefficient of the MS qCEDM operator is obtained from the matrix element with external quark-antiquark-gluon states:
\begin{align}
	\int d^Dx e^{-i q \cdot x} & \< \psi(k) | \O_{CE}^R(x;t) | \psi(p) g(r) \> \big|_\text{amp} \nn
		&= (2\pi)^D \delta^{(D)}(p+r-k-q) \< \psi(k) | \O_{CE}^R(0;t) | \psi(p) g(r) \> \big|_\text{amp} \nn
		&=: (2\pi)^D \delta^{(D)}(p+r-k-q) \epsilon_\mu^a(r) \M^a_\mu(p,k,r) \, .
\end{align}
The list of Feynman diagrams is shown in Fig.~\ref{img:cEDMDiagrams}. There are 19 additional diagrams, which follow from crossing the quark- and antiquark legs and inverting the fermion-flow direction. We also do not show the diagrams needed for the calculation of the quark-field renormalization. For $\xi=\alpha_0=1$, several diagrams are of second order in the soft scales and can be discarded, but they contribute in the calculation with generic gauge parameters.

In total, we obtain for $q=0$ (i.e., without additional momentum insertion into the operator)
\begin{align}
	\M_\mu^a(p,k,r) = -2i \tilde\sigma_{\mu\nu} r_\nu t^a Z_\chi^{-1} \zeta_\chi^{-1} + \frac{\alpha_s t^a}{4\pi} &\bigg[  -\frac{3}{4} \left( \frac{1}{\varepsilon} + \log(8\pi\mu^2 t) \right) (C_A - 8 C_F) \gamma_5 (k_\mu + p_\mu) \nn
		& + \frac{i}{4} \left( \frac{1}{\varepsilon} + \log(8\pi\mu^2 t) \right) (13C_A + 8 C_F) \tilde\sigma_{\mu\nu} r_\nu \nn
		& - \frac{1}{8} \Big((5+4\delta_\mathrm{HV})C_A + (68-48\delta_\mathrm{HV})C_F\Big) \gamma_5 (k_\mu + p_\mu) \nn
		& + \frac{i}{8}\Big((27+36\delta_\mathrm{HV})C_A-(44-16\delta_\mathrm{HV})C_F\Big) \tilde\sigma_{\mu\nu} r_\nu \bigg] \, .
\end{align}
Up to the contribution of equation-of-motion operators, this results in the following matching coefficient, including the finite renormalization imposed by~\eqref{eq:vevRenormalizationCondition}:
\begin{align}
	c_{CE}(t,\mu) &= \zeta_\chi^{-1} + \frac{\alpha_s}{4\pi} \left[ 2(C_F - C_A) \log(8\pi\mu^2 t) - \frac{1}{2} \Big( (4+5 \delta_\mathrm{HV}) C_A + (3- 4\delta_\mathrm{HV} ) C_F \Big) \right] \nn
		&= 1 + \frac{\alpha_s}{4\pi} \begin{aligned}[t]
			&\biggl[ (5C_F - 2C_A) \log(8\pi\mu^2 t) \nn
			&- \frac{1}{2} \Big( (4+5 \delta_\mathrm{HV}) C_A + (3- 4\delta_\mathrm{HV} ) C_F \Big) - \log(432) C_F \biggr] \, . \end{aligned} \mytag
\end{align}
The divergences of the expanded flowed diagrams cancel in the matching equations against the counterterms on the MS side, which are determined by the anomalous dimension of the qCEDM operator. We have again checked that the result for the matching coefficient is independent of the gauge parameters $\xi$ and $\alpha_0$.

\subsection{Chromo-MDM}
\label{sec:qCMDM}

Similarly to the qCEDM operator, we also define the CP-even flowed qCMDM operator in terms of renormalized (ringed) fields as
\begin{align}
	\O_{CM}^R(x;t) = \mathring{\bar\chi}(x;t) \sigma_{\mu\nu} t^a \mathring\chi(x;t) G_{\mu\nu}^a(x;t) \, .
\end{align}
Its short flow-time OPE up to dimension five reads
\begin{align}
	\O_{CM}^R(x;t) &= c_m(t,\mu) \O_{m}^\mathrm{MS}(\mu) + c_{m^3}(t,\mu) \O_{m^3}^\mathrm{MS}(\mu) + c_{m^5}(t,\mu) \O_{m^5}^\mathrm{MS}(\mu) \nn
		&\quad + c_S(t,\mu) \O_S^\mathrm{MS}(x;\mu) + c_{m^2S}(t,\mu) \O_{m^2S}^\mathrm{MS}(x;\mu) + c_{mG}(t,\mu) \O_{mG}^\mathrm{MS}(x;\mu) \nn
		&\quad + c_M(t,\mu) \O_M^\mathrm{MS}(x;\mu) + c_{CM}(t,\mu) \O_{CM}^\mathrm{MS}(x;\mu) + \ldots \, ,
\end{align}
where the renormalized MS operators are the minimally subtracted versions of
\begin{align}
	\O_m(\mu) &= \mu^{-2\varepsilon} m \, , \nn
	\O_{m^3}(\mu) &= \mu^{-2\varepsilon} m^3 \, , \nn
	\O_{m^5}(\mu) &= \mu^{-2\varepsilon} m^5 \, , \nn
	\O_S(x;\mu) &= \bar\psi(x) \psi(x) \, , \nn
	\O_{m^2S}(x;\mu) &= m^2 \bar\psi(x) \psi(x) \, , \nn
	\O_{mG}(x;\mu) &= \mu^{-2\varepsilon} m\, \tr[ G_{\mu\nu} G_{\mu\nu} ] \, , \nn
	\O_M(x;\mu) &= \bar\psi(x) \sigma_{\mu\nu} F_{\mu\nu}(x) \psi(x)  \, , \nn
	\O_{CM}(x;\mu) &= \bar\psi(x) \sigma_{\mu\nu} t^a \psi(x) G_{\mu\nu}^a(x) \, .
\end{align}

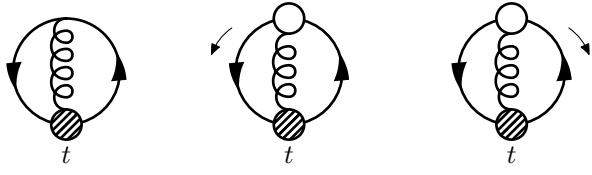
\begin{figure}[t]
	\centering
	\begin{fmfgraph*}(60,60)
		\fmfbottom{b1} \fmftop{t1} \fmfleft{l1} \fmfright{r1}
		\fmf{phantom,tension=8}{b1,v1}
		\fmf{phantom,tension=8}{t1,v2}
		\fmf{quark,right}{v1,v2,v1}
		\fmffreeze
		\fmf{gluon}{v2,v1}
		\fmfv{decor.shape=circle, decor.filled=shaded, decor.size=(4mm),label.angle=-90,label.dist=3mm,label=$t$}{v1}
	\end{fmfgraph*}
	\qquad
	\begin{fmfgraph*}(60,60)
		\fmfbottom{b1} \fmftop{t1} \fmfleft{l1} \fmfright{r1}
		\fmf{phantom,tension=8}{b1,v1}
		\fmf{phantom,tension=8}{t1,v2}
		\fmf{quark,right}{v1,v2,v1}
		\fmffreeze
		\fmf{marrowl,right}{v2,v1}
		\fmf{gluon}{v2,v1}
		\fmfv{decor.shape=circle, decor.filled=shaded, decor.size=(4mm),label.angle=-90,label.dist=3mm,label=$t$}{v1}
		\fmfv{decor.shape=circle, decor.filled=empty, decor.size=(4mm)}{v2}
	\end{fmfgraph*}
	\qquad
	\begin{fmfgraph*}(60,60)
		\fmfbottom{b1} \fmftop{t1} \fmfleft{l1} \fmfright{r1}
		\fmf{phantom,tension=8}{b1,v1}
		\fmf{phantom,tension=8}{t1,v2}
		\fmf{quark,right}{v1,v2,v1}
		\fmffreeze
		\fmf{marrowr,left}{v2,v1}
		\fmf{gluon}{v2,v1}
		\fmfv{decor.shape=circle, decor.filled=shaded, decor.size=(4mm),label.angle=-90,label.dist=3mm,label=$t$}{v1}
		\fmfv{decor.shape=circle, decor.filled=empty, decor.size=(4mm)}{v2}
	\end{fmfgraph*}
	\caption{Feynman diagrams for the matching calculation of the flowed qCMDM operator onto the identity.}
	\label{img:cMDMVacuumDiagrams}
\end{figure}

Due to chiral symmetry, mixing with the identity or with the gluon kinetic term requires an insertion of a mass factor, which we include in the definition of the operators. The mixing with the identity starts at two loops and is determined by the vacuum diagrams shown in Fig.~\ref{img:cMDMVacuumDiagrams}: only the first diagram gives a non-vanishing contribution.

The remaining coefficients can be calculated in analogy to the CP-odd case. The results agree with the CP-odd sector for $\dHV=0$:
\begin{align}
	c_m(t,\mu) &= -\frac{24 i N_c}{(4\pi)^2} \frac{\alpha_s C_F}{4\pi} \frac{1}{t^2} \log\left( \frac{4}{3} \right) \, , \nn
	c_{m^3}(t,\mu) &= -\frac{24 i N_c}{(4\pi)^2} \frac{\alpha_s C_F}{4\pi} \frac{1}{t} \left[ 2 \log(8\pi\mu^2 t) + \frac{1}{3} + 6 \log(2) - 3 \log(3) \right] \, , \nn
	c_{m^5}(t,\mu) &= -\frac{24 i N_c}{(4\pi)^2} \frac{\alpha_s C_F}{4\pi} \begin{aligned}[t]
		&\bigg[ 2 \log^2(8\pi\mu^2 t) + \frac{37}{6} \log(8\pi\mu^2 t) + \frac{\pi^2}{6} + \frac{1}{2} - 4 \log^2(2) + \frac{21}{2} \log(2) \nn
		& - \frac{21}{4} \log(3) + 2 \log(2)\log(3) - \mathrm{Li}_2\left( \frac{3}{4} \right) \bigg] \, , \end{aligned} \nn
	c_S(t,\mu) &= \frac{\alpha_s C_F}{4\pi} \frac{6i}{t} \, , \nn
	c_{m^2S}(t,\mu) &  = \frac{\alpha_s C_F}{4\pi} i \left[  12 \log(8\pi\mu^2 t) + \frac{29}{2} \right] \, , \nn
	c_{mG}(t,\mu) &= -\frac{i}{4\pi^2} \left[ 1 + \log(8\pi\mu^2t) \right] \, , \nn
	c_M(t,\mu) &= \frac{\alpha_s C_F}{4\pi} \left[ 4 \log(8\pi\mu^2 t) + 3 \right] \, , \nn
	c_{CM}(t,\mu) &= \zeta_\chi^{-1} + \frac{\alpha_s}{4\pi} \left[ (2C_F - 2C_A) \log(8\pi\mu^2 t) - \frac{1}{2} \Big( 4 C_A + 3 C_F \Big) \right] \nn
		&= 1 + \frac{\alpha_s}{4\pi} \biggl[ (5C_F - 2C_A) \log(8\pi\mu^2 t) - \frac{1}{2} \Big( 4 C_A + 3 C_F \Big) - \log(432) C_F \biggr] \, . 
\end{align}

In the case of the qCMDM, flavor off-diagonal components are also of interest \cite{Constantinou:2015ela,Constantinou:2017sgv},
because they can mediate  BSM contributions to  $K\rightarrow \pi\pi$ and  $\varepsilon^\prime/\varepsilon$. The matching coefficients 
$c_S(t,\mu)$, $c_M(t,\mu)$ and $c_{CM}(t,\mu)$ are the same for flavor diagonal and off-diagonal components, while only the diagonal components contribute to $c_{mG}(t,\mu)$, $c_{m}$, $c_{m^3}$ and $c_{m^5}$.
For the flavor-changing components of the qCMDM, the factor $m^2$ in $\O_{m^2 S}$ is replaced by $m^2 \mapsto (m_i^2 + m_f^2)/2$, with $i$ and $f$ the flavors of initial- and final-state quarks. With this replacement, $c_{m^2 S}$ is unchanged.


\section{Scale dependence of the matching coefficients}
\label{sec:scale}

\subsection{Scale dependence and perturbative uncertainty}

The SFTE connects renormalized operators at positive and 
vanishing flow time $t$. Operators at positive flow time defined in terms of the flowed gauge field and the ringed quark fields
are independent of the renormalization scale $\mu$, which implies that the scale dependence 
of the matching coefficients has to be cancelled by the renormalization-scale dependence of the renormalized MS operator at vanishing flow time,
up to higher-order corrections in the perturbative expansion.

In Sec.~\ref{sec:ren}, we have defined appropriately renormalized flowed operators and determined the matching coefficients in the SFTE. The matching coefficients depend on the matching scale and the flow time in the combination $\log(8\pi\mu^2 t)$. The additional scale dependence of the coupling $\alpha_s$ is beyond the accuracy of the one-loop matching calculation.

\begin{figure}[t]
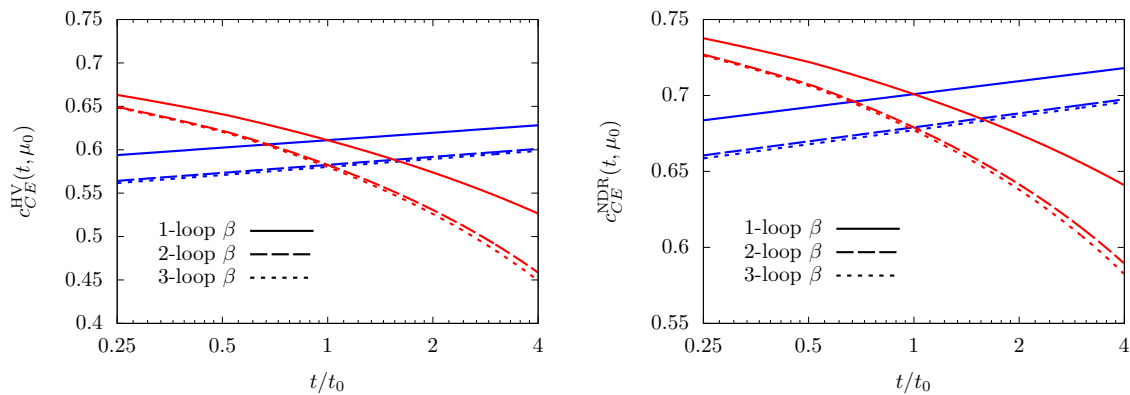

	\centering
	\scalebox{0.8}{
		\input{plots/matchingHV}
		\input{plots/matchingNDR}}
	\caption{Scale dependence of the matching coefficient $c_{CE}$. The left (right) plot shows the result in the HV (NDR) scheme. Detailed explanations are given in the main text.}
	\label{img:ScaleDependence}
\end{figure}

For illustration, we numerically evaluate the matching coefficient $c_{CE}(t,\mu)$. We take as input the \msbar{} coupling at the weak scale $\alpha_s(M_Z^2) = 0.1179$~\cite{ParticleDataGroup:2020ssz} and evolve it down to an \msbar{} scale $\bar\mu = 3\GeV$ using the one-, two-, or three-loop QCD $\beta$-function~\cite{Tarasov:1980au,Larin:1993tp}. This corresponds to an MS scale of
\begin{align}
	\mu_0 = 3\GeV \times \frac{e^{\gamma_E/2}}{(4\pi)^{1/2}} \approx 1.13\GeV \, .
\end{align}
In Fig.~\ref{img:ScaleDependence}, we plot the results for the matching coefficient $c_{CE}(t,\mu_0)$, evaluated for different values of the flow-time $t$ around
\begin{align}
	t_0 = \frac{1}{8\pi\mu_0^2} \, .
\end{align}
The deviation from the tree-level result $c_{CE} = 1$ illustrates the impact of the one-loop corrections, which are larger in the HV scheme than in NDR. The blue curves show the results for $\alpha_s$ evaluated at the fixed scale $\mu_0$, while the red curves show the result for $\alpha_s$ evaluated at a MS scale $\mu = (8\pi t)^{-1/2}$. In principle, we could resum the leading logarithms $(\alpha_s \log(8\pi\mu^2 t) )^n$ in the matching coefficient by solving the renormalization-group equations. However, the logarithms in the matching are small---for a matching performed at the flow time $t_0$, they vanish. We checked that the resummation has a small impact in the range of $t$ shown in Fig.~\ref{img:ScaleDependence}.

The mild logarithmic $t$-dependence of the blue curves simply reflects the scale dependence of the MS operators as dictated by the one-loop anomalous dimensions. The red curves lead to a result that differs from the blue curves by
\begin{align}
	\Delta = A_1 \alpha_s^2(\mu_0^2) \log^2(8\pi\mu_0^2 t) + A_2 \alpha_s^2(\mu_0^2) \log(8\pi\mu_0^2 t) + \O(\alpha_s^3) \, ,
\end{align}
with constant coefficients $A_i$ and where the subleading logarithm dominates numerically and is beyond the accuracy of our calculation. We take the maximal difference between the blue and red curves in the range $t\in[t_0/4, 4t_0]$ as an estimate of genuine $\O(\alpha_s^2)$ corrections, which require a two-loop matching calculation: this points to a relative uncertainty at these scales of about $\sim 10\%$--$20\%$.

\subsection{Non-perturbative effects and scheme dependence}

A comment is in order about perturbative estimates of power divergences in the flow time.
We have calculated the matching coefficients including the ones involving lower-dimensional operators: the coefficients $c_P(t,\mu)$, $c_m(t,\mu)$, $c_{m^3}(t,\mu)$, and $c_S(t,\mu)$ are power divergent for vanishing flow time.
Similarly to what happens with power divergences with the lattice spacing~\cite{Maiani:1991az}, perturbative estimates of these matching coefficients are not sufficient to enable a reliable extraction of the matrix elements of the MS dipole operators. Non-perturbative effects leave unsubtracted divergent or finite
terms. In Ref.~\cite{Kim:2021qae}, a first non-perturbative estimate of the power divergent
mixing of the qCEDM operator with the pseudoscalar density has been provided, and a smooth transition, toward small coupling,
to the perturbative result for $c_P(t,\mu) \sim t^{-1}$ obtained in Ref.~\cite{Rizik:2020naq} has been observed. 
The perturbative estimates of $c_P(t,\mu)$, $c_m(t,\mu)$, $c_{m^3}(t,\mu)$, and $c_S(t,\mu)$
obtained in this work thus provide a useful guide, at small coupling, for a non-perturbative 
determination of the same coefficients.

A possible strategy to determine the renormalized MS dipole operators at vanishing flow time
is to define the flowed dipole operators in terms of the ringed fields as in Eq.~\eqref{eq:qCEDMFlowed}
and then to provide a non-perturbative lattice definition of the matching coefficient 
of the power-divergent term. In this way the matching coefficient will depend on the scheme 
defined by the ringed fields, as well as on the renormalization 
scheme used to renormalize the lower-dimensional operator at vanishing flow time. However, since this operator renormalizes multiplicatively, the dependence on the scheme at vanishing flow time cancels in the product of the matching coefficient and operator matrix element. Therefore, the scheme dependence of the power divergence is only through the renormalization scheme at finite flow time (e.g., the definition of the ringed fields), which can be chosen in a regularization-independent way and subtracted non-perturbatively.
Once the non-perturbative subtraction of the power divergence is performed, in the continuum limit,
the subtracted operator
\begin{align}
	\O_{\textrm{sub}}(x;t) = \O^R_{CE}(x;t) - c_P^S(t,\mu) \O_P^S(x;\mu) \,,
\end{align}
where the superscript $S$ denotes the non-perturbative scheme at zero flow time (which can be defined even in a renormalization-group invariant way), will now have a SFTE with contributions only from operators 
of the same or higher dimensions.
The expansion can then be studied using standard techniques. In particular the target qCEDM operator at vanishing flow time needs to be renormalized in the scheme used to determine the matching coefficients. Similar subtractions can be applied to the CP-even operators.


\section{Summary and outlook}
\label{sec:summary}

The next generation of experimental searches for a permanent nucleon electric dipole moment (EDM) are expected to improve the precision of current constraints by an order of magnitude or more. The neutron EDM offers a unique window into  new sources of charge and parity (CP) violation generated by phenomena beyond the standard model of particle physics. The effects of heavy BSM particles on measurements at hadronic scales can be parametrized by effective, higher-dimensional CP-violating operators. In order to constrain the coefficients of these effective operators from data and ultimately obtain constraints on BSM theories, one needs to know the nonperturbative hadronic matrix elements of the effective operators.

We have calculated the short flow-time expansion coefficients for the quark chromo-EDM and the quark chromo-MDM operators. We have determined the complete set of one-loop matching coefficients up to dimension five, extending the work on the leading coefficients in Ref.~\cite{Rizik:2020naq}. These finite one-loop matching corrections enter at next-to-leading-log accuracy. For the qCMDM operator, we also included the two-loop matching to the identity operator.

This work is part of ongoing efforts to calculate the hadronic matrix elements of CP-violating operators with lattice QCD by the SymLat collaboration~\cite{Shindler:2014oha,Shindler:2015aqa,Dragos:2018uzd,Dragos:2019oxn} and others~\cite{Shintani:2005xg,Berruto:2005hg,Aoki:2008gv,Guo:2015tla,Alexandrou:2015spa,Abramczyk:2017oxr,Yoon:2017tag}. Our calculation provides the matching relations necessary to relate results extracted from lattice QCD to the MS (or \msbar{}) scheme, which is the one used in perturbative effective-field theory calculations. They are important to obtain robust constraints on BSM operator coefficients from the phenomenological analysis of experimental data. In addition, the perturbative calculation of the short flow-time coefficients will help to constrain the nonperturbative determination of the matching coefficient, first carried out in~\cite{Kim:2021qae}, by constraining it in the weak-coupling regime.

\begin{acknowledgments}
We thank V.~Cirigliano and J.~de Vries for useful discussions and collaboration at an early stage of the project. 
We further acknowledge several interesting discussions with T.~Bhattacharya, R.~Harlander, A.~Hasenfratz,
S.~Mondal, and O.~Witzel.
EM is supported by the USDOE through the Los Alamos National Laboratory and by the Laboratory Directed Research and Development program of Los Alamos National Laboratory under project number 20190041DR.
Los Alamos National Laboratory is operated by Triad National Security, LLC, for the National Nuclear Security Administration of U.S. Department of Energy (Contract No. 89233218CNA000001). 
CJM is supported in part by USDOE grant No.~DE-AC05-06OR23177, under which Jefferson Science Associates, LLC, manages and operates Jefferson Lab.
MDR and AS acknowledge funding support under the National Science Foundation grant PHY-1913287.
PS gratefully acknowledges financial support by the Swiss National Science Foundation (Project No.~PCEFP2\_194272).
\end{acknowledgments}

\appendix


\section{Conventions}
\label{sec:conventions}

\subsection[$SU(3)$]{\boldmath $SU(3)$}
We use traceless and anti-Hermitian $SU(3)$ generators $t^a$,
\begin{align}
	t^a = -i \frac{\lambda^a}{2} \, ,
\end{align}
where $\lambda^a$ are the Gell-Mann matrices. The generators fulfill
\begin{align}
	[ t^a, t^b ] = f^{abc} t^c \, , \quad  \{ t^a, t^b \} = - \frac{1}{3} \delta^{ab} - i d^{abc} t^c \, , \quad \tr[t^a t^b] = -\frac{1}{2} \delta^{ab} \, , \quad t^a t^a = - C_F \, .
\end{align}

\subsection{Dirac algebra and dimensional regularization}

We use the Dirac algebra in $D = 4 - 2\varepsilon$ Euclidean dimensions either in the HV or NDR scheme:
\begin{align}
	\{ \gamma_\mu, \gamma_\nu \} = 2 \delta_{\mu\nu} \, .
\end{align}
The Dirac matrices are Hermitian, $\gamma_\mu^\dagger = \gamma_\mu$ and fulfill
\begin{align}
	\gamma_\mu \gamma_\mu = \delta_{\mu\mu} = D \, .
\end{align}

In the HV scheme, we decompose the metric tensor into a part projecting onto $4$ and $-2\varepsilon$ dimensions, respectively,
\begin{align}
	\delta_{\mu\nu} = \bar \delta_{\mu\nu} + \hat \delta_{\mu\nu} \, ,
\end{align}
satisfying
\begin{align}
	\bar \delta_{\mu\nu} \bar \delta_{\nu\rho} &= \bar \delta_{\mu\rho} \, , \quad \hat \delta_{\mu\nu} \hat \delta_{\nu\rho} = \hat \delta_{\mu\rho} \, , \quad \bar \delta_{\mu\nu} \hat \delta_{\nu\rho} = 0 \, , \quad \bar \delta_{\mu\nu} \bar \delta_{\nu\mu} = 4 \, , \quad \hat \delta_{\mu\nu} \hat \delta_{\nu\mu} = -2\varepsilon \, .
\end{align}
The projections of gamma matrices (or four-vectors in general) are defined as
\begin{align}
	\bar\gamma_\mu = \bar\delta_{\mu\nu} \gamma_\nu \, , \quad \hat \gamma_\mu = \hat \delta_{\mu\nu} \gamma_\nu \, .
\end{align}
The fifth gamma matrix is defined as
\begin{align}
	\gamma_5 = \frac{1}{4!} \epsilon_{\mu\nu\rho\sigma} \gamma_\mu\gamma_\nu\gamma_\rho\gamma_\sigma = \gamma_1 \gamma_2 \gamma_3 \gamma_4 \, ,
\end{align}
with the purely four-dimensional antisymmetric Levi-Civita tensor, $\epsilon_{1234} = +1$. The fifth gamma matrix is Hermitian, $\gamma_5^\dagger = \gamma_5$, it anticommutes with the four-dimensional gamma matrices, and it commutes with the gamma matrices in the $D-4$-dimensional subspace:
\begin{align}
	\{ \gamma_5, \bar \gamma_\mu \} = [ \gamma_5 , \hat \gamma_\mu ] = 0 \, .
\end{align}
The HV scheme is algebraically consistent, and since QCD is a vector theory, spurious anomalies that arise in chiral gauge theories are absent in the present context (as long as we only consider single-operator insertions).

In the NDR scheme, $\gamma_5$ is assumed to anticommute with all Dirac matrices in $D$ dimensions.

In four dimensions, the Dirac matrices fulfill the Chisholm identity
\begin{align}
	\bar\gamma_\alpha \bar\gamma_\beta \bar\gamma_\gamma = \bar\gamma_\alpha \bar\delta_{\beta\gamma} + \bar\gamma_\gamma \bar\delta_{\alpha\beta} - \bar\gamma_\beta \bar\delta_{\alpha\gamma} - \bar\gamma_\mu \gamma_5 \epsilon_{\alpha\beta\gamma\mu} \, .
\end{align}


\section{Feynman rules}
\label{sec:FeynmanRules}

We start from QCD in Euclidean space in $D=4-2\varepsilon$ dimensions. We absorb the gauge couplings into the gauge fields and treat the electromagnetic field as an external static field.
The Euclidean QCD Lagrangian is given by
\begin{align}
	\L_\mathrm{QCD+GF+gh} &= \frac{1}{4 g_0^2} G_{\mu\nu}^a G_{\mu\nu}^a + \bar\psi ( \slashed D + m ) \psi + \L_\mathrm{GF} + \L_\mathrm{gh} \, , \nn
	\L_\mathrm{GF} &= \frac{1}{2 g_0^2 \xi} (\p_\mu G_\mu^a)^2 \, , \nn
	\L_\mathrm{gh} &= (\p_\mu \bar c^a) D_\mu^{ac} c^c \, ,
\end{align}
where the covariant derivative is
\begin{align}
	D_\mu &= \p_\mu + G_\mu + A_\mu \, , \quad G_\mu = t^a G_\mu^a \,,
\end{align}
when acting on fields which take values in the fundamental representation of $SU(N)$, or
\begin{align}
	D_\mu ( \cdot ) &= \p_\mu ( \cdot ) + [ G_\mu , \,\cdot\, ] \, , \quad D_\mu^{ac} = \p_\mu \delta^{ac} + f^{abc} G^b_\mu \,,
\end{align}
when acting on objects in the adjoint representation. The field-strength tensors are related to the commutator of the covariant derivative by
\begin{align}
	[ D_\mu, D_\nu ] &= G_{\mu\nu} + F_{\mu\nu} \, , \nn
	G_{\mu\nu} &= \p_\mu G_\nu - \p_\nu G_\mu + [ G_\mu, G_\nu ] \, , \nn
	F_{\mu\nu} &= \p_\mu A_\nu - \p_\nu A_\mu \, .
\end{align}
The Feynman rules are obtained from the generating functional
\begin{align}
	Z_E[J] = \int \mathcal{D}G \, \mathcal{D}\bar\psi \, \mathcal{D}\psi \, \mathcal{D}\bar c \, \mathcal{D}c \, e^{-S_E[J]} \, ,
\end{align}
where the Euclidean action including sources $J = \{ J_\mu^a, \zeta, \bar\zeta, N^a, \bar N^a\}$ is defined as
\begin{align}
	S_E[J] = \int \dD x \, ( \L_\mathrm{QCD+GF+gh} - J_\mu^a G_\mu^a - \bar\psi \zeta - \bar \zeta \psi - \bar c^a N^a - \bar N^a c^a) \, .
\end{align}
The standard QCD interaction vertices are
\begin{align}
	\begin{gathered}
		\begin{fmfgraph*}(60,50)
			\fmftop{t1} \fmfbottom{b1,b2}
			\fmf{quark,tension=2}{b1,v1}
			\fmf{quark,tension=2}{v1,b2}
			\fmf{gluon,label=$\mu,,a\;$,tension=3}{t1,v1}
		\end{fmfgraph*}
	\end{gathered}
	\quad &= - \gamma_\mu t^a \, , \nn
	\begin{gathered}
		\begin{fmfgraph*}(60,50)
			\fmftop{t1} \fmfbottom{b1,b2}
			\fmf{ghost,label=$c$,tension=2}{b1,v1}
			\fmf{ghost,label=$p,,a$,label.side=left,tension=2}{v1,b2}
			\fmf{gluon,label=$\mu,,b\;$,tension=3}{t1,v1}
		\end{fmfgraph*}
	\end{gathered}
	\quad &= i p_\mu f^{abc} \, , \text{ (for $p$ outgoing)}\, , \nn[0.3cm]
	\begin{gathered}
		\begin{fmfgraph*}(60,50)
			\fmftop{t1} \fmfbottom{b1,b2}
			\fmf{gluon,label=$p_1,,\mu,,a$,label.side=left,tension=3}{t1,v1}
			\fmf{gluon,label=$p_2,,\nu,,b\quad$,label.dist=-25,label.side=left,tension=2}{b1,v1}
			\fmf{gluon,label=$\quad p_3,,\rho,,c$,label.dist=-25,label.side=left,label.side=left,tension=2}{v1,b2}
		\end{fmfgraph*}
	\end{gathered}
	\qquad &=  -\frac{i f^{abc}}{g_0^2} \Big( \delta_{\mu\nu} ( p_1 - p_2)_\rho + \delta_{\nu\rho} ( p_2 - p_3 )_\mu + \delta_{\mu\rho} ( p_3 - p_1 )_\nu \Big) \, , \nn*
		&\quad \text{(all momenta outgoing)} \, , \nn[0.5cm]
	\begin{gathered}
		\begin{fmfgraph*}(60,50)
			\fmftop{t1,t2} \fmfbottom{b1,b2}
			\fmf{gluon,label=$p_1,,\mu,,a\quad$,label.dist=-20,tension=3}{t1,v1}
			\fmf{gluon,label=$\quad p_2,,\nu,,b$,label.dist=-23,label.side=right,tension=3}{v1,t2}
			\fmf{gluon,label=$p_3,,\rho,,c\qquad$,label.dist=-30,label.side=left,tension=2}{b1,v1}
			\fmf{gluon,label=$\qquad p_4,,\sigma,,d$,label.dist=-30,label.side=left,tension=2}{v1,b2}
		\end{fmfgraph*}
	\end{gathered}
	\qquad &= - \frac{1}{g_0^2} \begin{aligned}[t] &\Big( f^{ade} f^{bce} \left(\delta_{\mu \nu } \delta_{\rho \sigma }-\delta_{\mu \rho } \delta_{\nu \sigma }\right) \nn
		& + f^{ace} f^{bde} \left(\delta_{\mu \nu } \delta_{\rho\sigma }-\delta_{\mu \sigma } \delta_{\nu \rho }\right) \nn
		& + f^{abe} f^{cde} \left(\delta_{\mu \rho } \delta_{\nu \sigma }-\delta_{\mu \sigma } \delta_{\nu \rho}\right) \Big)  \, . \end{aligned} \mytag
\end{align}

The leading-order flowed propagators are given by:
\begin{align}
	s,\nu,b \;
	\begin{gathered}
		\begin{fmfgraph*}(60,50)
			\fmfleft{l1} \fmfright{r1}
			\fmf{gluon,label=$p$,label.dist=-15}{l1,r1}
		\end{fmfgraph*}
	\end{gathered} \;
	t,\mu,a \;
	&= \tilde D_{\mu\nu}^{ab}(p,s+t) = \int \dD x e^{-i p \cdot x} \wick{ \c B_\mu^a(x;t) \c B_\nu^b(0;s) } \nn
		&= g_0^2 \delta^{ab} \frac{1}{p^2} \left[ \left( \delta_{\mu\nu} - \frac{p_\mu p_\nu}{p^2} \right) e^{-(s+t)p^2} + \xi \frac{p_\mu p_\nu}{p^2} e^{-\alpha_0 (s+t) p^2} \right] \, , \\
	s,\beta \;
	\begin{gathered}
		\begin{fmfgraph*}(60,50)
			\fmfleft{l1} \fmfright{r1}
			\fmf{quark,label=$p$,label.dist=-15}{l1,r1}
		\end{fmfgraph*}
	\end{gathered} \;
	t,\alpha \;
	 &= \tilde S^{\alpha\beta}(p,s+t) = \int \dD x e^{-i p \cdot x} \wick{ \c \chi^\alpha(x;t) \c{\bar\chi}^\beta(0;s) } \nn
		&= \delta^{\alpha\beta} \frac{-i \slashed p + m}{p^2+m^2} e^{-(s+t)p^2} \, .
	\label{eq:QuarkPropagator}
\end{align}
At second and third order, instead of contracting the fundamental fields, one can reinsert the leading-order solution of the flow equation into the flow equation itself and replace the gauge field by second- and third-order expressions in the fields. The kernels of the flow equations then act as another type of ``propagator,'' and the interaction terms in~\eqref{eq:FlowInteractionTerms} generate interaction vertices with three and four fields. The same results are obtained by considering the $D+1$-dimensional field theory and calculating contractions with the Lagrange-multiplier fields~\cite{Artz:2019bpr}. We regard the flow-time integrals as part of the vertices and keep Heaviside step functions in the flow lines:
\begin{align}
	s,\nu,b \;
	\begin{gathered}
		\begin{fmfgraph*}(60,50)
			\fmfleft{l1} \fmfright{r1}
			\fmf{gluon,label=$p$,label.dist=-15}{l1,r1}
			\fmf{marrowd,tension=0}{l1,r1}
		\end{fmfgraph*}
	\end{gathered} \;
	t,\mu,a \;
	&= \delta^{ab} \theta(t-s) \int \dD x e^{-i p\cdot x} K_{\mu\nu}(x;t-s) \nn
	&= \delta^{ab} \theta(t-s) \frac{1}{p^2} \left[ (\delta_{\mu\nu}p^2 - p_\mu p_\nu) e^{-(t-s)p^2} + p_\mu p_\nu e^{-\alpha_0 (t-s) p^2} \right] \, ,
\end{align}
where the adjacent arrow points into the direction of increasing flow time.\footnote{Different diagrammatic notations can be found in the literature: in~\cite{Makino:2014taa,Rizik:2020naq}, flow lines are denoted by double lines. Here, we largely follow the convention of~\cite{Artz:2019bpr}.} For the quarks, one obtains
\begin{align}
	s,\beta \;
	\begin{gathered}
		\begin{fmfgraph*}(60,50)
			\fmfleft{l1} \fmfright{r1}
			\fmf{quark,label=$p$,label.dist=-15}{l1,r1}
			\fmf{marrowd,tension=0}{l1,r1}
		\end{fmfgraph*}
	\end{gathered} \;
	t,\alpha \;
	&= \delta^{\alpha\beta} \theta(t-s) \int \dD x e^{-i p\cdot x} J(x;t-s)
	= \delta^{\alpha\beta} \theta(t-s) e^{-(t-s)p^2} \, , \nn
	s,\beta \;
	\begin{gathered}
		\begin{fmfgraph*}(60,50)
			\fmfleft{l1} \fmfright{r1}
			\fmf{quark,label=$p$,label.dist=10}{r1,l1}
			\fmf{marrowd,tension=0}{l1,r1}
		\end{fmfgraph*}
	\end{gathered} \;
	t,\alpha \;
	&= \delta^{\alpha\beta} \theta(t-s) \int \dD x e^{-i p\cdot x} \bar J(x;t-s)
	= \delta^{\alpha\beta} \theta(t-s) e^{-(t-s)p^2} \, .
\end{align}
The gauge flow-vertices are easily obtained by considering Green's functions of three and four fields and replacing one field by the leading-order solution of the flow equation. The contractions of the fields then lead to one flow line and two or three flowed propagators, times the interaction vertex. The vertex rules are given by
\begin{align}
	\nn
	\begin{gathered}
		\begin{fmfgraph*}(70,60)
			\fmfleft{l1,l2} \fmfright{r1}
			\fmf{gluon,label=$\hspace{-1.5cm}p_3,,\rho,,c$,label.side=right,label.dist=35,tension=2}{l1,v1}
			\fmf{gluon,label=$\hspace{-1.5cm}p_2,,\nu,,b\quad\quad$,label.dist=-50,label.side=left,tension=2}{v1,l2}
			\fmf{gluon,label=$\quad\; t\quad p_1,,\mu,,a$,label.dist=5,label.side=right,label.side=left,tension=2}{v1,r1}
			\fmfv{decor.shape=circle, decor.filled=empty, decor.size=(4mm)}{v1}
			\fmf{marrowd,tension=0}{v1,r1}
			\fmf{darrowl,tension=0}{l1,v1}
			\fmf{darrowl,tension=0}{l2,v1}
		\end{fmfgraph*}
	\end{gathered}
	\quad &=  - i f^{abc} \int_0^\infty dt \, \begin{aligned}[t]
				& \Big( \delta_{\nu\rho} ( p_2 - p_3)_\mu + 2 \delta_{\mu\rho} {p_3}_\nu - 2 \delta_{\mu\nu} {p_2}_\rho \\
				&+ (\alpha_0-1) (  \delta_{\mu\nu} {p_3}_\rho - \delta_{\mu\rho} {p_2}_\nu ) \Big) \, , \end{aligned} \nn
		&\quad \text{(all momenta outgoing)} \, , \\[0.75cm]
	\begin{gathered}
		\begin{fmfgraph*}(80,70)
			\fmfleft{l1} \fmftop{t1} \fmfbottom{b1} \fmfright{r1}
			\fmf{gluon,label=$\Bigg._{\displaystyle p_4,,\sigma,,d}$,label.side=right,label.dist=10,tension=2}{v1,b1}
			\fmf{gluon,label=$\hspace{-1.5cm}p_3,,\rho,,c$,label.side=left,label.dist=5,tension=2}{l1,v1}
			\fmf{gluon,label=$\Bigg.^{\displaystyle p_2,,\nu,,b}$,label.dist=10,label.side=right,tension=2}{t1,v1}
			\fmf{gluon,label=$\quad\; t\quad p_1,,\mu,,a$,label.dist=5,label.side=right,label.side=left,tension=2}{v1,r1}
			\fmfv{decor.shape=circle, decor.filled=empty, decor.size=(4mm)}{v1}
			\fmf{marrowd,tension=0}{v1,r1}
			\fmf{darrowd,tension=0}{l1,v1}
			\fmf{darrowl,tension=0}{t1,v1}
			\fmf{darrowl,tension=0}{b1,v1}
		\end{fmfgraph*}
	\end{gathered}
	\quad &= - \int_0^\infty dt \, \begin{aligned}[t]
				& \Big( f^{abe} f^{cde} (\delta_{\mu\rho}\delta_{\nu\sigma} -  \delta_{\mu\sigma}\delta_{\rho\nu} )  \nn
				& + f^{ace} f^{bde} ( \delta_{\mu\nu}\delta_{\rho\sigma} - \delta_{\mu\sigma}\delta_{\nu\rho} )  \nn
				& + f^{ade} f^{bce} ( \delta_{\mu\nu}\delta_{\rho\sigma} - \delta_{\mu\rho}\delta_{\nu\sigma} )  \Big) \, , \end{aligned} \mytag
\end{align}
where the dashed adjacent arrows indicate that the line either is a propagator or a flow line, since in more complicated diagrams the leading-order solution of the flow equation can be reinserted iteratively. The fermionic interaction vertices are
\begin{align}
	\nn
	\begin{gathered}
		\begin{fmfgraph*}(70,60)
			\fmfleft{l1,l2} \fmfright{r1}
			\fmf{quark,label=$\hspace{-1.5cm}p_2$,label.side=right,label.dist=35,tension=2}{l1,v1}
			\fmf{gluon,label=$\hspace{-1.5cm}p_1,,\mu,,a\quad\quad$,label.dist=-50,label.side=left,tension=2}{v1,l2}
			\fmf{quark,label=$t\qquad$,label.dist=5,label.side=right,label.side=left,tension=2}{v1,r1}
			\fmfv{decor.shape=circle, decor.filled=empty, decor.size=(4mm)}{v1}
			\fmf{marrowd,tension=0}{v1,r1}
			\fmf{darrowl,tension=0}{l1,v1}
			\fmf{darrowl,tension=0}{l2,v1}
		\end{fmfgraph*}
	\end{gathered}
	\quad &=  - i t^a \int_0^\infty dt \, \begin{aligned}[t]
				& \Big( (1-\alpha_0) {p_1}_\mu + 2 {p_2}_\mu \Big) \, , \end{aligned} \nn
		&\quad \text{($p_1$, $p_2$ outgoing)} \, , \\[0.75cm]
	\begin{gathered}
		\begin{fmfgraph*}(70,60)
			\fmfleft{l1,l2} \fmfright{r1}
			\fmf{quark,tension=2}{l1,v1}
			\fmf{gluon,label=$\hspace{-1.5cm}p_1,,\mu,,a\quad\quad$,label.dist=-50,label.side=left,tension=2}{v1,l2}
			\fmf{quark,label=$t\qquad p_2$,label.dist=5,label.side=right,label.side=left,tension=2}{v1,r1}
			\fmfv{decor.shape=circle, decor.filled=empty, decor.size=(4mm)}{v1}
			\fmf{darrowd,tension=0}{r1,v1}
			\fmf{marrowl,tension=0}{v1,l1}
			\fmf{darrowl,tension=0}{l2,v1}
		\end{fmfgraph*}
	\end{gathered}
	\quad &=  i t^a \int_0^\infty dt \, \begin{aligned}[t]
				& \Big( (1-\alpha_0) {p_1}_\mu + 2 {p_2}_\mu \Big) \, , \end{aligned} \nn
		&\quad \text{($p_1$, $p_2$ outgoing)} \, , \\[0.75cm]
	\begin{gathered}
		\begin{fmfgraph*}(80,70)
			\fmfleft{l1} \fmftop{t1} \fmfbottom{b1} \fmfright{r1}
			\fmf{gluon,label=$\Bigg._{\displaystyle \nu,,b}$,label.side=right,label.dist=10,tension=2}{v1,b1}
			\fmf{quark,tension=2}{l1,v1}
			\fmf{gluon,label=$\Bigg.^{\displaystyle \mu,,a}$,label.dist=10,label.side=right,tension=2}{t1,v1}
			\fmf{quark,label=$\! t\qquad$,label.dist=5,label.side=right,label.side=left,tension=2}{v1,r1}
			\fmfv{decor.shape=circle, decor.filled=empty, decor.size=(4mm)}{v1}
			\fmf{marrowd,tension=0}{v1,r1}
			\fmf{darrowd,tension=0}{l1,v1}
			\fmf{darrowl,tension=0}{t1,v1}
			\fmf{darrowl,tension=0}{b1,v1}
		\end{fmfgraph*}
	\end{gathered}
	\quad &=  \delta_{\mu\nu}  \{ t^a, t^b \} \int_0^\infty dt \, , \\[0.75cm]
	\begin{gathered}
		\begin{fmfgraph*}(80,70)
			\fmfleft{l1} \fmftop{t1} \fmfbottom{b1} \fmfright{r1}
			\fmf{gluon,label=$\Bigg._{\displaystyle \nu,,b}$,label.side=right,label.dist=10,tension=2}{v1,b1}
			\fmf{quark,tension=2}{l1,v1}
			\fmf{gluon,label=$\Bigg.^{\displaystyle \mu,,a}$,label.dist=10,label.side=right,tension=2}{t1,v1}
			\fmf{quark,label=$\! t\qquad$,label.dist=5,label.side=right,label.side=left,tension=2}{v1,r1}
			\fmfv{decor.shape=circle, decor.filled=empty, decor.size=(4mm)}{v1}
			\fmf{darrowd,tension=0}{r1,v1}
			\fmf{marrowd,tension=0}{v1,l1}
			\fmf{darrowl,tension=0}{t1,v1}
			\fmf{darrowl,tension=0}{b1,v1}
		\end{fmfgraph*}
	\end{gathered}
	\quad &=  \delta_{\mu\nu}  \{ t^a, t^b \} \int_0^\infty dt \, .
\end{align}
The conventions for the vertex rules are chosen so that symmetry factors of loop diagrams match the ones of standard perturbation theory. Flow lines and propagators need to be distinguished when determining the symmetry factor of a given topology. E.g., comparing the two diagrams
\begin{align}
	\nn[1cm] &\mbox{\hspace{6.5cm}} , \\[-2cm]
	&\includegraphics[width=3cm]{images/diag-cEDM-9} \quad
	\includegraphics[width=3cm]{images/diag-cEDM-27} \nn[-1.25cm] \nonumber
\end{align}
the first one, as a usual QCD diagram, comes with a symmetry factor $1/S = 1/2$, while the second diagram involving one flow line and one propagator has a symmetry factor $1$.

Finally, we give the vertex rules for insertions of the effective operators. (Including an operator in the Euclidean Lagrangian $\L_\mathrm{eff} = \L + c_\O \O$ would result in $-c_\O$ times our vertex rules.) For the qCEDM, we obtain:
\begin{align}
	\nn
	\begin{gathered}
		\begin{fmfgraph*}(70,60)
			\fmfleft{l1,l2} \fmfright{r1}
			\fmf{quark,tension=2}{l1,v1}
			\fmf{gluon,label=$\hspace{-1.5cm}p,,\mu,,a\quad\quad$,label.dist=-50,label.side=left,tension=2}{v1,l2}
			\fmf{quark,label=$t\qquad$,label.dist=5,label.side=right,label.side=left,tension=2}{v1,r1}
			\fmfv{decor.shape=circle, decor.filled=shaded, decor.size=(4mm)}{v1}
			\fmf{darrowd,tension=0}{r1,v1}
			\fmf{darrowl,tension=0}{l1,v1}
			\fmf{darrowl,tension=0}{l2,v1}
		\end{fmfgraph*}
	\end{gathered}
	\quad &=  2i t^a \tilde\sigma_{\mu\nu}p_\nu \, , \quad \text{(for $p$ outgoing)} \, , \\[0.5cm]
	\begin{gathered}
		\begin{fmfgraph*}(80,70)
			\fmfleft{l1} \fmftop{t1} \fmfbottom{b1} \fmfright{r1}
			\fmf{gluon,label=$\Bigg._{\displaystyle \nu,,b}$,label.side=right,label.dist=10,tension=2}{v1,b1}
			\fmf{quark,tension=2}{l1,v1}
			\fmf{gluon,label=$\Bigg.^{\displaystyle \mu,,a}$,label.dist=10,label.side=right,tension=2}{t1,v1}
			\fmf{quark,label=$\! t\qquad$,label.dist=5,label.side=right,label.side=left,tension=2}{v1,r1}
			\fmfv{decor.shape=circle, decor.filled=shaded, decor.size=(4mm)}{v1}
			\fmf{darrowd,tension=0}{r1,v1}
			\fmf{darrowd,tension=0}{l1,v1}
			\fmf{darrowl,tension=0}{t1,v1}
			\fmf{darrowl,tension=0}{b1,v1}
		\end{fmfgraph*}
	\end{gathered}
	\quad &= 2 \tilde\sigma_{\mu\nu} f^{abc} t^c \, ,
\end{align}
while the vertex rules for the qCMDM are obtained by replacing $\tilde\sigma_{\mu\nu} \mapsto \sigma_{\mu\nu}$. In the matching equation, we also require the tree-level matrix elements of the MS operators. The vertex rule for the qEDM operator reads:
\begin{align}
	\nn[-0.25cm]
	\begin{gathered}
		\begin{fmfgraph*}(70,60)
			\fmfleft{l1,l2} \fmfright{r1}
			\fmf{quark,tension=2}{l1,v1}
			\fmf{photon,tension=2}{v1,l2}
			\fmf{quark,tension=2}{v1,r1}
			\fmflabel{$p,\mu$}{l2}
			\fmfv{decor.shape=circle, decor.filled=shaded, decor.size=(4mm)}{v1}
		\end{fmfgraph*}
	\end{gathered}
	\quad &=  2i \tilde\sigma_{\mu\nu}p_\nu \, , \quad \text{(for $p$ outgoing)} \, ,
\end{align}
while the qMDM vertex rules is obtained by replacing $\tilde\sigma_{\mu\nu} \mapsto \sigma_{\mu\nu}$. The two-gluon vertex rule for the QCD $\theta$ term $\O_{m\theta}$ reads
\begin{align}
	\begin{gathered}
		\begin{fmfgraph*}(70,30)
			\fmfleft{l1} \fmfright{r1}
			\fmf{gluon,label=$p_1,,\mu,,a\phantom{b}$,label.side=left,label.dist=5}{v1,l1}
			\fmf{gluon,label=$\;\;p_2,,\nu,,b$,label.dist=5,label.side=right,label.side=left}{r1,v1}
			\fmfv{decor.shape=circle, decor.filled=shaded, decor.size=(4mm)}{v1}
		\end{fmfgraph*}
	\end{gathered}
	\quad &= -2 m \delta^{ab} \epsilon_{\mu\nu\lambda\sigma} {p_1}_\lambda {p_2}_\sigma \, , \quad \text{(outgoing momenta)} \, ,
\end{align}
while for the gluon kinetic operator $\O_{mG}$ we obtain
\begin{align}
	\begin{gathered}
		\begin{fmfgraph*}(70,30)
			\fmfleft{l1} \fmfright{r1}
			\fmf{gluon,label=$p_1,,\mu,,a\phantom{b}$,label.side=left,label.dist=5}{v1,l1}
			\fmf{gluon,label=$\;\;p_2,,\nu,,b$,label.dist=5,label.side=right,label.side=left}{r1,v1}
			\fmfv{decor.shape=circle, decor.filled=shaded, decor.size=(4mm)}{v1}
		\end{fmfgraph*}
	\end{gathered}
	\quad &= 2 m \delta^{ab} \left( \delta_{\mu\nu} p_1 \cdot p_2 - {p_2}_\mu {p_1}_\nu \right) \, , \quad \text{(outgoing momenta)} \, .
\end{align}
The rule for quark bilinear operators $\bar\psi(x) \Gamma \psi(x)$ (scalar or pseudoscalar density) simply reads
\begin{align}
	\begin{gathered}
		\begin{fmfgraph*}(70,30)
			\fmfleft{l1} \fmfright{r1}
			\fmf{quark}{l1,v1}
			\fmf{quark}{v1,r1}
			\fmfv{decor.shape=circle, decor.filled=shaded, decor.size=(4mm)}{v1}
		\end{fmfgraph*}
	\end{gathered}
	\quad &=  \Gamma \, .
\end{align}

\end{myfmf}

\bibliographystyle{utphysmod}
\bibliography{qcedm_pt.bib}

\end{document}